\documentclass[twocolumn,a4paper,reprint,amssymb,aps,prd,groupedaddress,superscriptaddress,preprintnumbers,amsmath,nobibnotes,showpacs,nofootinbib]{revtex4-2}
\pdfoutput=1
\usepackage{graphicx}
\usepackage{float}
\usepackage{epsf}
\usepackage{bm}

\usepackage{amsmath}
\usepackage{mathtools}
\usepackage[T1]{fontenc}
\usepackage{amsfonts}
\usepackage{booktabs}
\usepackage{amssymb}
\usepackage{epstopdf}
\usepackage{natbib}
\usepackage{color}
\usepackage[dvipsnames]{xcolor}
\usepackage{verbatim}
\usepackage{multirow}
\usepackage{physics}
\usepackage{pifont}
\usepackage{comment}
\usepackage{bm}
\usepackage{microtype}
\usepackage[colorlinks=true, linkcolor=blue , citecolor=blue, urlcolor=blue]{hyperref}

\usepackage{amsmath}
\usepackage[capitalize]{cleveref}
\usepackage[normalem]{ulem}
\usepackage{enumitem}
\usepackage{lipsum}
\usepackage{epstopdf}

\DeclareUnicodeCharacter{2212}{-}
\newcommand{\bea}{\begin{eqnarray*}}
\newcommand{\eea}{\end{eqnarray*}}
\newcommand{\be}{\begin{equation}}
\newcommand{\ee}{\end{equation}}
\newcommand{\ba}{\begin{eqnarray}}
\newcommand{\ea}{\end{eqnarray}}

\begin{document}

\title{Dark Energy with Constant Inertial Mass Density: Updated Constraints and Curvature-Induced Sign Transitions in $\rho_{\rm DE}$ and $\rho_{\rm DE}+p_{\rm DE}$}

\author{Luis A. Escamilla}
\email{torresl@itu.edu.tr}
\affiliation{Department of Physics, Istanbul Technical University, Maslak 34469 Istanbul, Türkiye}

\author{Berat Karadavut}
\email{bkaradavut@cse.yeditepe.edu.tr}
\affiliation{Department of Computer Engineering, Yeditepe University Ataşehir, 34755 Istanbul, Türkiye}

\author{Nihan Kat{\i}rc{\i}}
\email{nkatirci@dogus.edu.tr}
\affiliation{Department of Electrical-Electronics Engineering Doğuş University, \" Umraniye 34775 Istanbul, Türkiye}

\begin{abstract}

We present updated observational constraints on the simple-gDE model, characterized by a constant inertial mass density (IMD) $\rho_{\rm DE}+p_{\rm DE}$—which belongs to the broader graduated dark energy family—and compare its cosmological implications with those of the $w$CDM (constant $p_{\rm DE}/\rho_{\rm DE}$) and the $\Lambda$CDM (constant $\rho_{\rm DE}$) models. This parametrization provides a physically motivated, one-parameter extension of $\Lambda$CDM, perspective on dark-energy dynamics   beyond the usual equation-of-state approach. We use the newly released DESI DR2 Baryon Acoustic Oscillation (BAO) data in combination with either Cosmic Microwave Background (CMB) measurements from Planck 2018 or late-time probes, namely cosmic chronometers (CC) and the Pantheon+ Type Ia supernova (SNe Ia) sample, considered both with and without SH0ES calibration in this analysis. The data favor a small positive IMD, and Bayesian evidence indicates that the Simple-gDE, $w$CDM, and $\Lambda$CDM models remain statistically indistinguishable within spatially flat scenarios. Consequently, none of these models exhibits a sign transition in the dark-energy density, and no improvement in $H_0$ tension. Allowing spatial curvature qualitatively enlarges the phenomenology of the dark sector. In particular, the interplay between spatial curvature and a nonzero inertial mass density permits sign transitions in both the effective dark-energy density and the IMD during cosmic evolution. For the BAO+CC+SN+SH0ES dataset, the $o$Simple-gDE model yields a transition redshift $z^\dagger = 1.51^{+0.68}_{-0.34}$ in the $W_{-1}$ branch, while the crossing of the Null Energy Condition boundary (NECB), defined by $\rho_{\rm DE}+p_{\rm DE}=0$, occurs at 
$z_{\rm NECB}=2.36^{+1.48}_{-1.48}$. The model is statistically favored over $o\Lambda$CDM and $ow$CDM according to Bayesian evidence. These results highlight the potential role of inertial mass density as a fundamental parameter in dark-energy phenomenology and demonstrate that geometric effects, such as spatial curvature, can reveal dynamical features of the dark sector that remain hidden within the spatially flat $\Lambda$CDM framework. 
\end{abstract}

\maketitle

\section{Introduction}

The current cosmological paradigm is described by the standard model of cosmology, $\Lambda$CDM, where the cosmological constant ($\Lambda$) and Cold Dark Matter (CDM) account for dark energy (DE) and dark matter (DM), the two dominant components of the universe. Despite its simplicity and remarkable success in explaining a wide range of observations, a persistent discrepancy has emerged between direct measurements of the present expansion rate of the universe and those inferred from the early universe within the $\Lambda$CDM framework.

Cepheid-calibrated Type Ia supernova (SN Ia) measurements by the SH0ES collaboration yield 
$H_0 \simeq 73 \,\mathrm{km\, s^{-1}\, Mpc^{-1}}$~\cite{Riess:2021jrx,Breuval:2024lsv}, 
while cosmic microwave background (CMB) analyses prefer significantly lower values 
\cite{Planck:2018vyg,Planck:2018nkj,AtacamaCosmologyTelescope:2025blo,SPT-3G:2025bzu}. 
The latest combined Planck+SPT+ACT CMB analysis reports 
$H_0 = 67.24 \pm 0.35 \,\mathrm{km\, s^{-1}\, Mpc^{-1}}$~\cite{SPT-3G:2025bzu}. 
Comparing this with the local determination from the H0DN (“Local Distance Network”), 
$H_0 = 73.50 \pm 0.81 \,\mathrm{km\, s^{-1}\, Mpc^{-1}}$, 
leads to a discrepancy of $\simeq 7.1\sigma$~\cite{H0DN:2025lyy}. 
This result is based on a covariance-weighted synthesis of multiple local distance indicators, reducing dependence on any single calibration method \cite{Efstathiou:2020wxn,Mortsell:2021nzg,Mortsell:2021tcx,Riess:2021jrx,Sharon:2023ioz,Murakami:2023xuy,Riess:2023bfx,Brout:2023wol,Uddin:2023iob,Riess:2024ohe,Freedman:2024eph,Riess:2024vfa}. This disagreement, known as the Hubble tension, remains as one of the most significant challenges in modern cosmology (see~\cite{Verde:2019ivm,DiValentino:2020zio,DiValentino:2021izs,Perivolaropoulos:2021jda,Schoneberg:2021qvd,DiValentino:2022fjm,DiValentino:2024yew,Perivolaropoulos:2024yxv,Akarsu:2024qiq} for reviews). The absence of a fundamental understanding of the nature of DE and DM further complicates attempts to resolve this tension, motivating the present work. Beyond the Hubble tension, additional hints of possible departures from the $\Lambda$CDM framework have recently emerged. 
The Dark Energy Spectroscopic Instrument (DESI) 2024 Data Release 1 (DR1) BAO observations—both independently and in combination with CMB data~\cite{Planck:2018vyg} and Type Ia supernova measurements~\cite{Scolnic:2021amr,Brout:2022vxf,Rubin:2023ovl,DES:2024tys}—reported more than $2\sigma$ evidence for dynamical dark energy within the CPL framework \cite{DESI:2024mwx}. 
This indication has been further strengthened by the DESI Data Release 2 (DR2)~\cite{DESI:2025zgx}, with recent analyzes finding a $\sim 3\sigma$ preference for dynamical dark energy depending on the data combination and modeling assumptions \cite{DES:2025sig,Hoyt:2026fve}. The physical origin of this apparent dynamical behavior of dark energy remains unclear and may reflect modified gravity \cite{CANTATA:2021asi,CosmoVerse:2025txj}, alternative dark energy dynamics, or residual systematic effects in the observational data.
\cite{Efstathiou:2020wxn,Mortsell:2021nzg,Mortsell:2021tcx,Murakami:2023xuy,Sharon:2023ioz,Riess:2023bfx,Bhardwaj:2023mau,Uddin:2023iob,Brout:2023wol,Dwomoh:2023bro,Riess:2024ohe,Freedman:2024eph,Riess:2024vfa}.

Considering the energy conservation (continuity) equation,
$\dot{\rho} + 3H\rho(1+w) = 0$, 
the simplest phenomenological way to introduce dynamical dark energy beyond $\Lambda$—which corresponds to vacuum energy with constant density and equation of state (EoS) $w \equiv p/\rho = -1$—is to allow $w \neq -1$, assuming the dark energy density remains positive. In this case, the energy density evolves with the scale factor as $\rho(a)$ rather than remaining constant. A simple realization is the \textit{w}CDM model, where $w$ is constant but treated as a free parameter. A further extension allows a time-dependent EoS; the most widely used form is the $w_0w_a$ parametrization (Chevallier–Polarski–Linder, CPL) \cite{Chevallier:2000qy,Linder:2002et}, which represents $w(a)$ as a first-order Taylor expansion in $a$. See also Ref.~\cite{Ormondroyd:2025iaf} for a broader discussion of EoS parametrizations. However, these approaches implicitly assume that the dark energy density remains positive throughout cosmic history. Relaxing this assumption opens the possibility of richer dynamical behavior, including scenarios in which the dark energy density may approach or even cross zero during cosmic evolution. Recent findings have suggested the possibility of a sign transition in the dark energy density, where it adopts negative values in the early universe before transitioning to positive values at late times. 

Ref.~\cite{Akarsu:2019hmw} first proposed that the inertial mass density, $\varrho \equiv \rho+p$, may be more fundamental than the energy density and introduced a dark energy parametrization with a minimal dynamical deviation from vacuum energy ($\varrho=0$) in the form $\varrho \propto \rho^{\lambda}$, known as graduated dark energy (gDE) ($\varrho={\rm const.}$ corresponds to simple-gDE \cite{Acquaviva:2021jov}). For large negative $\lambda$, gDE suggests that around $z_{\dagger}\sim2$ the Universe may have experienced a rapid mirror AdS-to-dS transition in vacuum energy—a sign-switching cosmological constant from negative to positive while preserving its magnitude. The resulting $\Lambda_{\rm s}$CDM model ~\cite{Akarsu:2021fol,Akarsu:2022typ,Akarsu:2023mfb,Escamilla:2025imi},  has attracted attention because it can yield higher inferred values of $H_0$ and, in several analyzes, improved Bayesian evidence relative to the standard $\Lambda$CDM scenario, while also alleviating other mild tensions such as the $S_8$ discrepancy \cite{CosmoVerse:2025txj}.

Another important aspect is that the inertial mass density may, in fact, be more fundamental than the energy density itself. The success of the CPL parametrization largely stems from the fact that it allows the inertial mass density (IMD), $\varrho_{\rm CPL}=\rho(a)\big[1+w_0+w_a-w_a a\big]$, to change sign at the scale factor $a_{\rm NECB}=(1+w_0+w_a)/w_a$, corresponding to $z_{\rm NECB}\sim0.5$, where $z=-1+1/a$ denotes the redshift \cite{Ozulker:2025ehg}. Although this behavior is often referred to as ``phantom crossing,'' such a description is meaningful as long as the DE energy density remains positive, $\rho>0$, which is indeed the case for CPL. Throughout this work; we adopt the null energy condition boundary (NECB) definition, $\rho_{\rm DE}+p_{\rm DE}=0$, introduced in \cite{Akarsu:2026anp,Gokcen:2026pkq}. \cite{Caldwell:2025inn} clarifies that the Null Energy Condition applies to the total cosmological fluid rather than to individual components, implying that apparent violations inferred from the behavior of dark energy alone do not necessarily indicate a true NEC violation. Even scenarios where the effective dark energy satisfies negative inertial mass density and later crosses positive inertial mass density are viable without violating the NEC at the level of the total energy–momentum content of the universe.

In another successful model, the IMD vanishes in the abrupt $\Lambda_{\rm s}$CDM model. In contrast, $\Lambda_{\rm s}$CDM-type models~\cite{Alexandre:2023nmh,Akarsu:2025gwi,Akarsu:2025dmj,Akarsu:2024qsi,Akarsu:2024eoo,Akarsu:2024nas,Souza:2024qwd,DiGennaro:2022ykp,Nyergesy:2025lyi,Anchordoqui:2023woo,Anchordoqui:2024gfa,Anchordoqui:2024dqc,Soriano:2025gxd} allow the dark-energy density itself to change sign around $z\sim2$. It is therefore worthwhile to investigate whether the effective inertial mass density, $\rho_{\rm DE}+p_{\rm DE}$, may also undergo a sign transition in such scenarios.

However, despite its phenomenological success, the CPL parametrization cannot reproduce a DE energy density that crosses zero and have negative values in the past, and it limits its ability to capture such dynamics. In \cite{Gokcen:2026pkq}, inspired by \cite{Ozulker:2025ehg}, authors introduce two phenomenological extensions of the CPL framework that allow the dark energy density to change sign. BAO and SNeIa data push any negative-density phase beyond the effective redshift coverage of the observations, implying that such models are disfavored relative to CPL and that allowing $\rho_{\rm DE}<0$ reduces the statistical significance of the dynamical dark energy signal. In Ref.~\cite{Adil:2023exv}, authors defined the Omnipotent Dark Energy (ODE) class, characterized by its capacity to encompass all six combinations of the energy density sign ($\rho_{\rm DE} > 0$ or $\rho_{\rm DE} < 0$) and the EoS regime ($w_{\rm DE} > -1$, $w_{\rm DE} = -1$, or $w_{\rm DE} < -1$) within a single, unified cosmic history. 
Indeed, these combinations are physically governed by the inertial mass density, defined as $\varrho_{\rm DE} = \rho_{\rm DE} + p_{\rm DE} = \rho_{\rm DE}(1 + w_{\rm DE})$. In this context, the inertial mass density directly dictates the evolution of the energy density via the relation:
\begin{equation}
    \frac{d\rho_{\rm DE}}{dz} = \frac{3}{1+z}\varrho_{\rm DE}(z).
\end{equation}
 Under the assumption of a positive energy density, a positive inertial mass density ($\varrho_{\rm DE} > 0$) implies that the energy density decreases as the universe expands (or increases toward the past, i.e., higher $z$), effectively mimicking a quintessence-like scalar field. 
Conversely, an increase in dark energy density with redshift $z$ occurs when the inertial mass density is negative ($\varrho_{\rm DE} < 0$), characterizing a phantom-like behavior. However, the specific signature of the energy density evolution remains dependent on the individual values of $\rho_{\rm DE}$ and $p_{\rm DE}$. Within this framework, several distinct physical scenarios emerge: both positive and negative energy densities may either increase or decrease with redshift, depending on the interplay between the sign of $\rho_{\rm DE}$ and the regime of the equation of state.

For canonical scalar fields, the inertial mass density corresponds to the kinetic energy of the scalar field, $\rho_{\phi}+p_{\phi}=\dot{\phi}^{2}$,
while negative IMD formally corresponds to phantom-like scalar fields, $\dot{\phi}^{2}<0$. Current observational data are consistent with such a possibility. However, in a single canonical scalar-field model, the kinetic term has a fixed sign, implying that a sign transition in the inertial mass density cannot occur within a single-field canonical framework. Dataset combinations that favor a sign change in the IMD, can therefore be more naturally interpreted within the two-field (quintom) framework developed in \cite{Akarsu:2026anp}. For instance, in Horndeski \cite{Horndeski:1974wa,Deffayet:2011gz}, Kinetic Gravity Braiding (KGB) theories \cite{Deffayet:2010qz} and in Dirac–Born–Infeld (DBI) models, the relation $\rho+p \neq 2X P_X$ can arise due to derivative couplings. Such terms modify the effective inertial mass density and may allow negative IMD without introducing ghost instabilities. The possible ($\phi,\dot{\phi}$) combinations with $w\geq-1$ to those with $w<-1$ has been studied in \cite{Vikman:2004dc}. This is indeed the same situation achieved via $G_3$ and $G_4$ terms in Hordenski theories encountered in \cite{Tiwari:2023jle,Tiwari:2024gzo} and warrants further investigation. Remarkably, the possibility that simultaneous sign transitions in both $\rho_{\rm DE}$ and $\rho_{\rm DE}+p_{\rm DE}$ could help alleviate cosmological tensions appears to have gone largely noticed in the literature, yet \textit{subsequent sign transitions in first $\rho_{\rm DE}$ and then $\rho_{\rm DE}+p_{\rm DE}$ as we go today} behavior has not been well characterized before.

A previous joint analysis using Cosmic Chronometers (CC), Type Ia supernovae (SN), CMB (Planck 2018), and BAO (SDSS) data \cite{Acquaviva:2021jov} found nearly equivalent statistical evidence for both the $\Lambda$CDM model and DE with a constant inertial mass density scenario. The inferred value of the constant IMD was slightly positive, $\varrho_{\rm ci}=(3.06\pm2.28)\times10^{-31}\,{\rm g\,cm^{-3}}$ [$\mathcal{O}(10^{-12})\,\mathrm{eV}^4$]. This result points to a possible departure from the vanishing inertial mass density associated with the standard cosmological constant, favoring instead a small but non-zero dynamical quantity. This slightly positive value corresponds to a present-day equation-of-state parameter $w_{\rm ci0}=-0.948\pm0.041$ in the simple-gDE model, consistent with $w_0>-1$ in the CPL parametrization. However, resolving the $H_0$ tension typically requires phantom behavior ($w_{\rm DE}<-1$) in the past. Since a constant inertial mass density cannot generate such a NECB crossing or allow transitions to negative energy density regimes, its non-zero value requires further scrutiny with the latest observational datasets.

On the other hand, when spatial curvature is included, both possibilities may arise. In particular, the combined CC+SN+BAO dataset previously allowed notable deviations from $\Lambda$CDM within the simple-gDE framework: simple-gDE favored a spatially closed universe with $\Omega_{k0}=-0.122\pm0.117$, permitting a spatial curvature contribution of up to $\sim12\%$, while $\Lambda$CDM remained consistent with spatial flatness, but models had still shared the same evidence. We have not realized this structure, character of DE. In light of these recent findings and the potential sign transition of the inertial mass density indicated by the data, we update the observational constraints of \cite{Acquaviva:2021jov} using the newly released DESI DR2 BAO and Planck 2018 CMB measurements, together with the Pantheon+ Type Ia supernova sample with a model comparison  of the simple-gDE model with both the standard $\Lambda$CDM (constant vacuum energy density $\rho_{\rm vac}$) paradigm and the $w$CDM model. These two represent one-parameter extensions of this framework in which $w$ and $\rho+p$  are held constant. We further explore the role of spatial curvature, which effectively acts as a negative density component in a spatially closed universe. In particular, we examine the possibility of negative energy density transitions and potential crossings of the null energy condition boundary (NECB) arising from the interplay between spatial curvature and deviations from vanishing IMD in the dark energy sector.

In Sec.~\ref{sec:IMD}, we introduce the theoretical framework of dark energy with non-null inertial mass density and review the simple-gDE model characterized by a constant IMD. 
In Sec.~\ref{sec:curvatureDE}, we extend this framework by incorporating spatial curvature and discuss the conditions under which sign transitions in the dark-energy density and the inertial mass density may arise. 
Sec.~\ref{sec:obs} presents the methodology of the cosmological parameter estimation together with the observational datasets used in the analysis. 
In Sec.~\ref{sec:results}, we report the main results of the parameter constraints and perform a Bayesian comparison between the $\Lambda$CDM, $w$CDM, and simple-gDE models. Sec.~\ref{sec:negativeDE} explores the interplay between spatial curvature and non-null inertial mass density, focusing on the emergence of negative dark-energy density and possible NEC boundary crossings. 
Finally, we summarize our findings and discuss their implications in Sec.~\ref{sec:final}.

\section{Non-null DE Inertial mass density}
\label{sec:IMD}
Within general relativity, governed by the Einstein field equations, local conservation of energy and momentum is guaranteed as a consequence of the twice-contracted Bianchi identity,
\begin{align}
G_{\mu\nu}=-T_{\mu\nu} \;\;\Rightarrow\;\; \nabla_\mu G^{\mu\nu}=0 
\quad\text{and}\quad 
\nabla_\mu T^{\mu\nu}=0.
\end{align}

Projecting $\nabla_\mu T^{\mu\nu}=0$ parallel and orthogonal to $u^\mu$ yields the energy and momentum conservation equations, respectively,
\begin{eqnarray}
\label{eqn:constraint}
\dot{\rho}+\Theta\varrho=0,
\qquad
{\rm D}^\mu p+\varrho \dot{u}^\mu=0,
\end{eqnarray}
where $\Theta={\rm D}^\mu u_\mu$ denotes the volume expansion rate, a dot represents the derivative with respect to the comoving time $t$, and the relation $\nabla_\nu u_\mu={\rm D}_\nu u_\mu-\dot{u}_\mu u_\nu$ has been used \cite{Ellis:1971pg,Ellis:1998ct}. In the momentum conservation equation, ${\rm D}^\mu p$ is the pressure gradient, $\dot{u}^\mu$ is the four-acceleration, and 
\begin{eqnarray}
\varrho=\rho+p
\end{eqnarray}
defines the \textit{inertial mass density}, namely \textit{the enthalpy per unit volume}. Cosmologists commonly describe dark energy through equation-of-state parametrizations, $w=p/\rho$, such as the $w$CDM and CPL models being the most widely used. However, recent studies suggest that the dark energy density may have evolved from negative values in the early universe to positive values at late times
~\cite{Sahni:2002dx,Vazquez:2012ag,BOSS:2014hwf,Sahni:2014ooa,BOSS:2014hhw,DiValentino:2017rcr,Mortsell:2018mfj,Poulin:2018zxs,Capozziello:2018jya,Wang:2018fng,Banihashemi:2018oxo,Dutta:2018vmq,Banihashemi:2018has,Akarsu:2019ygx,Li:2019yem,Visinelli:2019qqu,Ye:2020btb,Perez:2020cwa,Akarsu:2020yqa,Calderon:2020hoc,Ye:2020oix,DeFelice:2020cpt,Paliathanasis:2020sfe,Bonilla:2020wbn,Acquaviva:2021jov,Bag:2021cqm,Bernardo:2021cxi,Escamilla:2021uoj,DiGennaro:2022ykp,Akarsu:2022lhx,Bernardo:2022pyz,Ong:2022wrs,Tiwari:2023jle,Malekjani:2023ple,Alexandre:2023nmh,Gomez-Valent:2023uof,Medel-Esquivel:2023nov,Anchordoqui:2023woo,Anchordoqui:2024gfa,Gomez-Valent:2024tdb,Bousis:2024rnb,Manoharan:2024thb,Wang:2024hwd,Colgain:2024ksa,Tyagi:2024cqp,Yadav:2024duq,Toda:2024ncp,Dwivedi:2024okk,Anchordoqui:2024dqc,Gomez-Valent:2024ejh,Souza:2024qwd,Mukherjee:2025myk,Giare:2025pzu,Keeley:2025stf,Soriano:2025gxd,Wang:2025dtk,Bouhmadi-Lopez:2025ggl,Bouhmadi-Lopez:2025spo,Paraskevas:2024ytz,Akarsu:2025ijk,Efstratiou:2025xou,Gonzalez-Fuentes:2025lei,Hogas:2025ahb,Mishra:2025goj}. Such scenarios inevitably lead to a divergence in the EoS when the energy density approaches zero ($\rho \to 0$). This singular behavior cannot be properly captured by standard $w$-based parametrizations, which assume smooth and bounded functions. Moreover, from the standpoint of parameter inference and numerical stability, these divergences pose serious challenges, since most likelihood estimators and numerical solvers are not designed to handle the infinite values implied by such EoS behavior.
At this juncture, Refs.~\cite{Akarsu:2019hmw,Acquaviva:2021jov} proposed that a deeper understanding of the nature of vacuum energy may require shifting the focus towards more fundamental physical quantities, such as the IMD of DE.
The energy density or the equation-of-state parameter alone may not provide sufficient information about a fluid's dynamics, whereas the inertial mass density offers greater insight. The covariant analogue of Newton’s second law,
\begin{align}
\nabla^{\mu} P = -(\rho + p)\,\dot{u}^{\mu} \equiv -\varrho\,\dot{u}^{\mu},
\end{align}
remains valid even when the IMD does not vanish for vacuum energy. This is because, in a homogeneous and isotropic universe, the four-acceleration of matter vanishes identically in the absence of external forces, ensuring that the momentum conservation equation is automatically satisfied regardless of the value of $\varrho$. 

In this study, we investigate a one-parameter extension of the $\Lambda$CDM model based on the IMD of DE, adopting a more fundamental physical perspective. We evaluate this approach against both the standard $\Lambda$CDM paradigm and its one-parameter extension, the $w$CDM model. Then we incorporate spatial curvature as a component of DE to the analysis to increase its capability for all three models. 

\subsection{$w$CDM: Constant EoS parameter having dynamical non-null IMD}

The $w$CDM model assumes a constant equation-of-state parameter for dark energy, leading to the well known evolution for the energy density:
\begin{equation}
\rho_{w{\rm CDM}}(z) = \rho_{\rm DE,0} (1+z)^{3(1+w)} \quad \textnormal{where} \quad w = \text{const}.
\end{equation}
In this framework, the inertial mass density evolves exactly the same with the energy density, yet scaled by the factor $(1+w)$:
\begin{align}
\varrho_{w{\rm CDM}}(z) &= \rho_{\rm DE,0} (1+w) (1+z)^{3(1+w)} \nonumber \\
&= (1+w)\rho_{w{\rm CDM}}(z),
\end{align}
whose sign is determined entirely by the factor $(1+w)$. Near the cosmological constant limit ($w \simeq -1$), the inertial mass density approaches zero ($\varrho_{w{\rm CDM}} \simeq 0$), while the energy density remains nearly constant, as it also appears in the exponent of $(1+z)$. We also recall that cosmic acceleration occurs when
$\rho + 3p < 0$, equivalent to the condition $w < -1/3$ for $\rho>0$.

\subsection{Simple Graduated Dark Energy: Constant IMD}
 Indeed, the simplest phenomenological generalization of the usual vacuum energy is then to promote its \textit{null inertial mass density} to an arbitrary constant~\cite{Acquaviva:2021jov}
\begin{equation}
\label{eqn:GDE}
\varrho=\rm const=\rho_{\rm ci}+p_{\rm ci}=\varrho_{\rm ci}, \quad \quad \textnormal{(simple-gDE)}
\end{equation}
for which the energy density $\rho_{\rm ci}$ (supposed to be positive today, i.e., $\rho_{\rm ci0}>0$) and the pressure $p_{\rm ci}$ are not necessarily constant---here and in what follows the subscript $0$ attached to any quantity denotes its present-day ($z=0$) value and subscript ``ci'' denotes \textit{constant inertia}. It is worth noting that this promotion corresponds to a possible suggestion of the IMD, instead of standard vacuum energy density, $\Lambda$, as one of the constants of nature.
Substituting \eqref{eqn:GDE} in \eqref{eqn:constraint}, the energy density and pressure of simple-gDE read: 
\begin{align}
\label{eqn:rhogde}
\rho_{\rm ci}(z)=&\rho_{\rm ci0}+3\varrho_{\rm ci} \ln (1+z),\\
p_{\rm ci}(z)=&-\rho_{\rm ci0}+\varrho_{\rm ci} \left[1-3\ln (1+z)\right],
\end{align}
 and the EoS parameter\footnote{The functional form and redshift dependence of the EoS resemble the phenomenological LOG parametrization in $w(a)$ form listed in Table~II of Ref.~\cite{DESI:2025fii}. Despite this similarity, our model contains one fewer parameter and exhibits fundamentally different dynamical behavior,} as follows:
\begin{equation}
\label{eq:eos}
  w_{\rm ci}(z)=-1+\frac{1+w_{\rm ci0}}{1+3\left(1+w_{\rm ci0}\right)\ln(1+z)}.
\end{equation}
Using geometric units $\hbar=c=8\pi G=1$, the Friedmann equation can be written as:
\begin{align}
\label{eq:fried}
\frac{H^2}{H_0^2} =\,
&\Omega_{\rm ci0}\left[1+3(1+w_{\rm ci0})\ln(1+z)\right]+\Omega_{\rm m0}(1+z)^3\nonumber \\
&+\Omega_{\rm r0}(1+z)^4 ,
\end{align}
here $\Omega_i=\rho_i/\rho_{\rm c}$ denotes the density parameter of the $i$-th component with $\rho_{\rm c}=3H^2$ being the critical energy density. We exclude the possibility $\rho_{\rm ci0}<0$, as it is manifestly incompatible with observational constraints. 

The model approaches the $\Lambda$CDM limit at early times since the cosmic dynamics are dominated by pressureless matter. In this regime, the logarithmic term— independent of the signature of $\varrho_{\rm ci}$—behaves asymptotically like a cosmological constant, yielding $w_{\rm ci}\simeq -1$ for $z\gg1$. On the other hand, the future evolution is directly determined by the sign of $\varrho_{\rm ci}$, as this component eventually dominates the total energy density. In particular, if $\varrho_{\rm ci}>0$, the Universe reaches a future bounce at a finite cosmic time whereas for $\varrho_{\rm ci}<0$, the expansion culminates in a Little Sibling of the Big Rip (LSBR) singularity in the infinite future~\cite{Bouhmadi-Lopez:2014cca}. For the latter case, unlike conventional phantom dark energy models characterized by $w_{\rm ci0}<-1$ and $\rho_{\rm ci0}>0$, this scenario displays two distinctive features:
\textbf{(i)} The energy density does not asymptotically vanish as $z$ increases; instead, it crosses zero at
\begin{equation}
\label{eq:zpole}
z_{\rm ci\dagger}=-1+e^{-\frac{1}{3(1+w_{\rm ci0})}},
\end{equation}
after which it continues to evolve toward increasingly negative values (often referred to as $n$-phantom, where $n$ denotes negative energy density), \textbf{(ii)} The effective EoS given in \eqref{eq:eos} satisfies $w_{\rm ci}<-1$ for $z<z_{\rm ci\dagger}$ and $w_{\rm ci}>-1$ for $z>z_{\rm ci\dagger}$. Provided that $w_{\rm ci0}\neq -1$, the EoS approaches $w_{\rm ci}\to -1$ both in the asymptotic future ($z\to -1$) and in the very early Universe ($z\to\infty$), while developing a pole at $z=z_{\rm ci\dagger}$ corresponding to the vanishing of the energy density. This pole lies in the finite past for $w_{\rm ci0}<-1$ and in the finite future for $w_{\rm ci0}>-1$. The limiting case $w_{\rm ci0}=-1$ reduces to standard vacuum energy, yielding $z_{\rm ci\dagger}=-1$ or $z_{\rm ci\dagger}=\infty$, which implies that no such transition occurs.

If the inertial mass density of vacuum energy is allowed to be negative in observational analysis, and if negative values are favored, its energy density decreases with redshift in a manner reminiscent of phantom models; yet it crucially differs by crossing zero at a finite redshift. Remarkably, a combined analysis of Hubble parameter, supernovae, Planck 2018 CMB, and SDSS BAO data performed in Ref.~\cite{Acquaviva:2021jov} found comparable statistical support for $\Lambda$CDM and for a dark energy component with constant inertial mass density. The inferred value,
\[
\varrho_{\rm ci}=(3.06\pm2.28)\times10^{-31}\,\mathrm{g\,cm^{-3}}
\sim\mathcal{O}(10^{-12})\,\mathrm{eV}^4 ,
\]
was slightly positive, favoring a small but nonzero deviation from the standard assumption of vanishing inertial mass density. Confirmation of this result with updated observational datasets remains an important open task. 

Here, we aim to demonstrate that a non-null inertial mass density is favored by multiple recent updated datasets, and to test both its constancy and its signature. We confront this scenario with one-parameter dark energy extensions and perform a statistical comparison among the models.

\section{Spatial curvature together with non-null constant IMD of DE}
\label{sec:curvatureDE}
It was previously shown in \cite{Acquaviva:2021jov} that spatial curvature alone can mimic a dynamical vacuum energy, effectively acting as a negative energy component in closed geometries, $\Omega_{k0}<0$ and contributing to late-time cosmic acceleration. 

The spatial curvature of the RW metric can effectively be treated as a source described by an EoS parameter $w_k=-1/3$ and the corresponding energy density reads
\begin{equation}
\label{eqn:ksource}
    \rho_{k}=\rho_{k0}(1+z)^2,
\end{equation}
for which $\rho_{k0}>0$, $\rho_{k0}=0$, and $\rho_{k0}<0$ correspond to spatially open, flat, and closed universes, respectively. 

Taking into account that the inferred inertial mass density remains slightly positive when treated as a non-zero constant, and recognizing that the standard 
$\Lambda$CDM and $w$CDM models strictly forbid any sign change in the dark energy density, we are compelled to explore more general scenarios. In such extended frameworks, the dark energy density may effectively change sign—a possibility that appears to be supported by current observational evidence. For this reason, deviations from null inertial mass density and from spatial flatness should be constrained both jointly and independently. Likewise, departures from $w=-1$ and from spatial flatness must be examined and statistically compared with the standard $\Lambda$CDM scenario as well as with simple gDE models. 

\subsection{ $o\Lambda$CDM and $ow$CDM models}
We define the effective source ($ow$) made up of the constant EoS ($w$) and the spatial curvature ($k$), for which the energy density, $$\rho_{ow}\equiv \rho_{w}+\rho_{k}$$ reads
 \begin{align}
\label{eq:fried1}
\Omega_{ow}=&\,\Omega_{w0}(1+z)^{3(1+w)}+\Omega_{k0}(1+z)^2,
\end{align}
where $\Omega_{k0}=\rho_{k0}/\rho_{\rm c}=3H^2$ is the density parameter corresponding to spatial curvature, being the critical energy density and $\Omega_{k0}$ can be positive (open) and negative (closed) spatial sections.
\subsubsection{Sign transition in DE energy density }
It follows directly from Eq.~\eqref{eq:fried1} that a sign transition can in principle occur even within the $\Lambda$CDM and $w$CDM frameworks. Such a transition arises when the curvature and dark-energy contributions balance each other, leading to
\begin{equation}
\label{eq:owdagger}
z_{ow\dagger}=-1+\left(-\frac{\Omega_{k0}}{\Omega_{w0}}\right)^{\frac{1}{1+3w}},
\end{equation}
which is defined for $w\neq -1/3$ (the effective equation of state of spatial curvature) and is real only if $\Omega_{\rm k0}/\Omega_{w0}<0$, i.e., when $\Omega_{k0}$ and $\Omega_{w0}$ have opposite signs. In the particular case where the $\Lambda$ and spatial curvature jointly contribute to the effective dark-energy sector in Eq.~\eqref{eq:owdagger}, the total energy density crosses zero at a single redshift
\begin{equation}
z_{o\Lambda\dagger}=-1+\sqrt{-\frac{\Omega_{\Lambda}}{\Omega_{k0}}},
\end{equation}
provided that $\Omega_{k0}$ and $\Omega_{\Lambda}$ have opposite signs.

\subsubsection{Sign transition in DE IMD }

We consider the effective source composed of the constant--EoS component ($w$) and the spatial curvature ($k$),
\begin{equation}
\rho_{ow}=\rho_w+\rho_k, \qquad p_{ow}=p_w+p_k .
\end{equation}

For the constant--EoS fluid we have
$p_w=w\rho_w$ and the spatial curvature term scales as $a^{-2}$ and can be written as an effective fluid with equation of state
$w_k=-\frac{1}{3}$ and $p_k=w_k\rho_k=-\frac{1}{3}\rho_k$. Therefore, the IMD of the effective source scaled by $\rho_{c0}$  becomes 
\begin{equation}
\label{eq:imd}
\frac{\rho_{ow}+p_{ow}}{\rho_{c0}}
=(1+w)\Omega_{w0}(1+z)^{3(1+w)}
+\frac{2}{3}\Omega_{k0}(1+z)^2 
\end{equation}
using the redshift evolution. The NECB transition redshift,
\begin{equation}
z_{ow\rm NECB}=-1+
\left[
-\frac{2\,\Omega_{k0}}
{3(1+w)\Omega_{w0}}
\right]^{\frac{1}{1+3w}},
\qquad (w\neq -1/3).
\end{equation}

A real solution exists only if
\begin{equation}
\frac{\Omega_{k0}}{(1+w)\Omega_{w0}}<0 .
\end{equation}

In $\Lambda$CDM, \eqref{eq:imd} reads
\begin{equation}
\frac{\rho_{o\Lambda}+p_{o\Lambda}}{\rho_{c0}}
=
\frac23\,\Omega_{k0}(1+z)^2,
\end{equation}
which implies the inertial mass density does not change sign at any finite redshift; its sign is entirely determined by the sign of $\Omega_{k0}$.

\subsection{o-Simple gDE}
The Friedmann equation giving the complete description of the model under consideration here reads
\begin{align}
\label{eq:fried2}
\Omega_{\rm kci}=\,&\Omega_{\rm ci0}\left[1+3(1+w_{\rm ci0}) \ln (1+z) \right]+\Omega_{k0}(1+z)^2.
\end{align}

\subsubsection{Sign transition in DE energy density}
If $w_{\rm ci0}<-1$, DE energy density crosses zero at only one redshift
 \begin{align}
z_{k{\rm ci}\dagger} =-1+{\rm e}^{-\frac{1}{3(1+w_{\rm ci0})}-\frac{1}{2} W_0(x)},
\end{align}
on the other hand, if $w_{\rm ci0}>-1$, we might have two different redshift values  
\begin{align}
z_{k{\rm ci\dagger}} =
\begin{aligned}
&-1+{\rm e}^{-\frac{1}{3(1+w_{\rm ci0})}-\frac{1}{2} W_{-1}(x)},\\
&-1+{\rm e}^{-\frac{1}{3(1+w_{\rm ci0})}-\frac{1}{2} W_0(x)},
\end{aligned}
\label{eq:tildezpole}
\end{align}
for 
\begin{align}
x=\frac{\Omega_{k0}}{\Omega_{\rm ci0}} \frac{2}{3(1+w_{\rm ci0})} {\rm e}^{-\frac{2}{3(1+w_{\rm ci0})}},
\end{align}
and $W_{0}(x)$ and $W_{-1}(x)$ are the two real branches of the Lambert $W(x)$ function.

This result indicates that a sign change in the dark energy density does not necessarily require $w_{\rm ci0}<-1$. For a closed spatial geometry, the simple-gDE model allows the dark energy density to become effectively negative in the past.

\subsubsection{Sign transition in DE IMD }

Effective inertial mass density is as follows:
\begin{equation}
\frac{\rho_{\rm kci}+p_{\rm kci}}{\rho_{c0}}
=
(1+w_{\rm ci0})\,\Omega_{\rm ci0}
+\frac{2}{3}\Omega_{k0}(1+z)^2.
\end{equation}
which yields the NECB transition redshift as
\begin{equation}
z_{\rm kci,NECB}
=
-1+\sqrt{-\frac{3(1+w_{\rm ci0})\,\Omega_{\rm ci0}}{2\,\Omega_{k0}}}.
\end{equation}
A real solution requires
\begin{equation}
-\frac{(1+w_{\rm ci0})\,\Omega_{\rm ci0}}{\Omega_{k0}}>0,
\end{equation}
so the existence of a $\rho+p$ sign change depends only on the relative signs of
$(1+w_{\rm ci0})\Omega_{\rm ci0}$ and $\Omega_{k0}$.

For instance, for a spatially closed universe ($\Omega_{k0}<0$), the inertial mass density
$\rho_{\rm kci}+p_{\rm kci}$ can change sign only if the constant contribution
$(1+w_{\rm ci0})\Omega_{\rm ci0}$ and the curvature term have opposite signs, namely constant inertial mass density is positive $(w_{\rm ci0}>-1)$. Therefore, in a closed geometry the model allows a NECB transition in the past ($z_{\rm kci,NECB} >0$) of the inertial mass density when the constant
$\rm ci$ contribution dominates sufficiently over the curvature term.

Therefore, deviations from the cosmological constant and/or spatial flatness should be confronted with the observational data together and separately, and we should check whether the effective source can assume negative density values in the finite past. Yet, a conclusive answer as to whether the model under consideration here \eqref{eq:fried} does better than the $\Lambda$CDM model cannot be given unless we rigorously confront the model with the latest observational data.

\section{Observational Data Analysis}
\label{sec:obs}
\subsection{Methodology of Cosmological Data Analysis}
\label{sec:observations}

In this study, we perform a parameter estimation analysis to place observational constraints on the free parameters of three different constant-parameter approaches to dark energy. Specifically, we consider the standard $\Lambda$CDM paradigm, the $w$CDM model (featuring a constant equation-of-state parameter $w$), and the \text{simple-gDE} model, which is characterized by a constant inertial mass density $\varrho_{\rm ci}$. These models are initially analyzed within a spatially flat universe. Subsequently, we extend our investigation to their non-flat counterparts, denoted as $o\Lambda$CDM, $ow$CDM, and $o\text{simple-gDE}$, respectively, to assess the impact of including spatial curvature as an additional free parameter.
To this end, we implement the theoretical models in a \texttt{SimpleMC} code \cite{SimpleMC} (initially released in Ref. \cite{BOSS:2014hhw}) and use a modified version of Markov Chain Monte Carlo (MCMC) sampler to perform MCMC analyses\footnote{MCMC methods are central to modern Bayesian inference, as they allow efficient sampling from complex posterior distributions. 
Introduced by Metropolis et al. \cite{Metropolis1953} and later generalized by Hastings \cite{Hastings1970}, these techniques have become essential for exploring high-dimensional parameter spaces and computing marginal distributions that are otherwise intractable through direct numerical methods.}.

In our numerical implementation, the radiation density parameter $\Omega_{r0}$ is treated as a derived quantity rather than a free parameter. Following standard cosmological conventions, we fix the CMB temperature to $T_\gamma = 2.7255$~K and set the effective number of relativistic species to $N_{\rm eff} = 3.046$. The present-day radiation contribution is then internally computed as:
\begin{equation}
    \Omega_{r0} = \Omega_{\gamma0} \left[1+\frac{7}{8}\left(\frac{4}{11}\right)^{4/3} N_{\rm eff} \right],
\end{equation}
where the photon density $\Omega_{\gamma0}$ is determined by the physical constants ($K_B$, $h_P$, $c$) and the Hubble constant via:
\begin{equation}
    \Omega_{\gamma0} = \frac{8\pi^5 (K_B T_\gamma)^4}{15 h_P^3 c^5} \frac{8 \pi G}{3 H_0^2}.
\end{equation}
Substituting the relevant numerical values, the relationship simplifies to:
\begin{equation}
    \Omega_{r0} = 4.183 \times 10^{-5} \, h_0^{-2},
    \label{eq:Omega_rad}
\end{equation}
where $h_0 = H_0 / (100\,\textrm{km s}^{-1}\,\textrm{Mpc}^{-1})$. While $\Omega_{r0}$ has a negligible impact on the expansion history at the low redshifts primarily explored in this study, it is retained within the Friedmann equation to ensure the robustness of our background modeling. Consequently, although this calculation is performed internally by the code, we do not report $\Omega_{r0}$ in our main results as its posterior is entirely dictated by the distribution of $h_0$.

The parameters for our investigated models are constrained using flat (non-informative) priors, as summarized in Table~\ref{tab:model_posteriors}. We adopt a standard range for the matter density $\Omega_{\rm m0}$ and the reduced adimensional Hubble constant $h_0$, while the physical baryon density $\Omega_{\rm b0}h_0^2$ is centered around the narrow range favored by Big Bang Nucleosynthesis (BBN) and CMB observations. 
For the model extensions, the spatial curvature parameter $\Omega_{\rm k0}$ is varied within the interval $[-0.6, 0.6]$ only for the non-flat ($o$) variants, effectively allowing for both open and closed geometries. In the $w$CDM framework, the constant equation-of-state parameter $w_0$ is allowed to explore both the quintessential and phantom regimes. For the simple-gDE model, we vary the present-day equation-of-state value $w_{\rm ci0}$ within the range $[-1.5,0]$.
This parameter choice allows us to test the consistency of a non-zero, constant inertia against the standard null-inertia ($\Lambda$CDM) case.

\begin{table}[H]
\centering
\renewcommand{\arraystretch}{1.4}
\begin{tabular}{lll}
\hline
\textbf{Model} & \textbf{Parameter} & \textbf{Prior Range} \\
\hline
\multirow{4}{*}{Base Parameters} & $\Omega_{\rm m0}$ & $[0.05, 1.0]$ \\
                               & $\Omega_{\rm b0}h_0^2$ & $[0.02, 0.025]$ \\
                               & $h_0$ & $[0.4, 1.0]$ \\
                               & $\Omega_{\rm k0}$ (non-flat) & $[-0.6, 0.6]$ \\
\hline
$(o)w$CDM                      & $w_0$ & $[-3.0, 1.0]$ \\
\hline
$(o)$Simple-gDE                & $w_{\rm ci0}$ & $[-1.5, 0]$ \\
\hline
\end{tabular}
\caption{Flat prior ranges adopted for the cosmological parameter estimation across the $\Lambda$CDM, $w$CDM, and simple-gDE models, including their spatially curved $(o)$ counterparts.}
\label{tab:model_posteriors}
\end{table}

\subsection{Data sets}  
Our baseline datasets are as follows: \\

\textbf{BAO} : the most recent (DR2) Baryon Acoustic Oscillation (BAO) measurements taken from the Dark Energy Spectroscopic Instrument (DESI)  \cite{DESI:2025zgx,DESI:2025zpo, DESI:2025fii,DESI:2025qqy} of galaxies, quasars, and the Lyman-$\alpha$ forest. These measurements provide constraints on the transverse comoving distance $D_{\rm M}/r_{\rm d}$, the Hubble horizon $D_{\rm H}/r_{\rm d}$ and the angle-averaged distance $D_{\rm V}/r_{\rm d}$ where the distances are normalized by the comoving sound horizon $r_{\rm d}$ at the drag epoch. To improve the constraining power on the parameter space, we also include a big bang nucleosynthesis (BBN) prior on the baryon contribution \cite{Mossa:2020gjc}; this is required to fix the $r_{d}$, particularly if we do not use Planck information. \\

\textbf{SN} : Pantheon+ analysis of $1701$ light curves of $1550$ distinct Type Ia supernovae (SNe Ia) ranging in redshift from $z=0.001$ to $2.26$ \cite{Scolnic:2021amr,Brout:2022vxf}. \\

\textbf{SH0ES}: Gaussian prior of $H_0 = 73.04 \pm 1.04 \, \mathrm{km} \, \mathrm{s}^{-1}\, \mathrm{Mpc}^{-1}$ from the local distance ladder measurements of the SH0ES team, based on the calibration of the absolute magnitude of SN Ia with Cepheid variables~\citep{Riess:2021jrx}. 
For comparison, the analysis will also include instances without this SH0ES prior. 
\\

\textbf{CC}:  Cosmic chronometers directly trace the differential evolution of cosmic time with redshift, thereby providing model-independent measurements of the Hubble expansion rate. We use a compilation of 15 CC measurements~\cite{Moresco:2020fbm}, provided together with their full covariance matrix, which allows correlated statistical and systematic uncertainties to be consistently accounted for in the likelihood analysis.\\

\textbf{CMB}: In \texttt{SimpleMC} \cite{SimpleMC}, we can employ a compressed \textit{Planck--2018} CMB likelihood using the acoustic scale $\theta_*=r_{\rm d}/D_{A}(z_*)$ ($100\theta_*=1.041085$ with a precision of  0.03\%~\cite{Planck:2018vyg}), along with physical baryon and cold dark matter (plus baryons) density,  $\Omega_{\rm b0} h_0^2$ and $\Omega_{\rm cb} h_0^2$, providing a robust high-redshift constraint on the background expansion while avoiding a full perturbation analysis. For this reason, we perform our analysis both with and without the reduced Planck CMB (coming from $z_*$, the redshift of last scattering), allowing us to realize genuine late-time solutions to cosmological tensions from those that rely on relaxing early-Universe constraints.

\subsection{Bayesian Model Comparison}

The statistical analysis of our model set $\{\mathcal{M}_i\}$ is performed by constructing a global likelihood function. Assuming the datasets are independent, the total log-likelihood is the sum of the individual likelihoods, where the minimum chi-squared is defined as $\chi_{\rm min}^2 = -2\ln{\mathcal{L}_{\rm max}}$. For a given dataset, the $\chi^2$ is expressed as:
\begin{equation}
\label{eq:chisq}
\chi_{\rm data}^2 = (d_{\mathcal{M}} - d_{\rm obs})^T C^{-1}_{\rm data} (d_{\mathcal{M}} - d_{\rm obs}),
\end{equation}
where $d_{\mathcal{M}}$ represents the model predictions, $d_{\rm obs}$ the observables, and $C_{\rm data}$ the covariance matrix. While the difference in minimum chi-squared, $\Delta\chi_{\rm min}^2$, provides a straightforward comparison between models with the same number of degrees of freedom, it does not account for model complexity or prior volumes.

To robustly compare models of varying complexity, we perform a Bayesian model comparison. This approach relies on the Bayesian evidence ($\mathcal{Z}$), which is the integral of the likelihood $\mathcal{L}(\text{Data}|\theta, \mathcal{M})$ over the prior space $\pi(\theta|\mathcal{M})$:
\begin{equation}
\mathcal{Z} = \int \mathcal{L}(\text{Data}|\theta, \mathcal{M}) \pi(\theta|\mathcal{M}) d\theta.
\end{equation}
The evidence naturally implements Occam’s razor, penalizing models with excessive parameters or unnecessarily large prior volumes. We compute the evidence from our MCMC chains using the \texttt{MCEvidence} code~\cite{Heavens:2017hkr,MCEvidence}, and compare models $i$ and $j$ using the Bayes factor:
\begin{equation}
\label{eq:LnZ}
\Delta \ln{\mathcal{Z}_{ij}} \equiv \ln \mathcal{Z}_i - \ln \mathcal{Z}_j.
\end{equation}
To interpret $\Delta \ln{\mathcal{Z}_{ij}}$, we adopt the revised Jeffreys' scale~\cite{Kass:1995loi}, where $|\Delta \ln \mathcal{Z}| < 1$ is considered inconclusive, $1 < |\Delta \ln \mathcal{Z}| < 3$ indicates moderate support, and $|\Delta \ln \mathcal{Z}| > 3$ suggests strong evidence for one model over the other. Our posterior distributions are sampled using eight independent MCMC chains per case. Convergence of the MCMC chains is monitored via the Gelman-Rubin diagnostic $R-1$~\cite{Gelman:1992zz}, where we enforce a strict threshold of $R-1 < 0.01$ to ensure the stability and accuracy of our results. The resulting chains are analyzed and visualized with the \texttt{GetDist} package \cite{Lewis:2019xzd}.

\section{Results and Discussion }
\label{sec:results}
\begin{table*}[t!]
  \caption{Mean values and $68\%$ confidence-level uncertainties for the cosmological parameters in the $\Lambda$CDM, $w$CDM, and Simple gDE models constrained with the BAO+CC+SN and BAO+CMB+SN datasets. $\chi_{\rm min}^2=-2\ln{\mathcal{L}_{\rm max}}$ is used to compare best fit with respect to the $\Lambda$CDM model. The last rows contain the Bayesian evidence $\ln \mathcal{Z}$ and the relative Bayesian evidence with respect to the $\Lambda$CDM, i.e., $\Delta\ln \mathcal{Z}=\ln \mathcal{Z}_{{\Lambda}\rm CDM}-\ln \mathcal{Z}$. }
  \label{tab:noknoshoes}
    \begin{tabular}{lcccccc}
  	\hline
    \toprule
    \multicolumn{1}{l}{Dataset} & \multicolumn{3}{c}{\textbf{BAO+CC+SN}} & \multicolumn{3}{c}{\textbf{BAO+CMB+SN}} \\  \hline
& \textbf{$\Lambda$CDM} & \textbf{$w$CDM} & \textbf{Simple-gDE}
& \textbf{$\Lambda$CDM} & \textbf{$w$CDM} & \textbf{Simple-gDE} \\
   \hline 
      \midrule
      \vspace{0.1cm}
$\Omega_{\rm m0}$ & $0.305 \pm 0.008$ & $0.301 \pm 0.009$ & $0.301 \pm 0.008$ & $0.305 \pm 0.006$ & $0.305 \pm 0.006$ & $0.308 \pm 0.013$ \\ \vspace{0.1cm}
$\Omega_{\rm b0}h_0^2$ &$0.0223 \pm 0.0003$& $0.0223 \pm 0.0004$ & $0.0223 \pm 0.0003$ & $0.0225\pm 0.0001$ & $0.0224 \pm 0.0001$ & $0.0225 \pm 0.0002$  \\ \vspace{0.1cm}
$h_0$  & $0.697 \pm 0.011$ & $0.673 \pm 0.018$  & $0.669 \pm 0.021$ & $0.682 \pm 0.004$  &  $0.684 \pm 0.006$ & $ 0.679 \pm 0.013$  \vspace{0.1cm}\\
 $w_{\rm ci0}$  &  $-1$  & $ -0.931 \pm 0.043 $ &$ -0.914 \pm 0.058$ &$-1$ &$-0.955 \pm 0.027 $  &  $-0.951 \pm 0.086$       \\ 
\hline
\vspace{0.1cm}
 $\varrho_{\rm ci0}\times 10^{31}$ $[\rm g\, cm^{-3}]$  &   0 & $4.09 \pm 2.12$ & $4.85 \pm 3.12$ & $0$ & $2.51 \pm 1.34$ &  $2.45 \pm 2.66$\\ \vspace{0.1cm}
$\Omega_{\rm ci0}$ & $0.649\pm0.008$ & $0.651\pm0.008$  & $0.649\pm 0.009$ & $0.646\pm0.006$ &  $0.647\pm0.007$ & $0.643\pm0.019$ \\  \vspace{0.10cm} 
 $z_{\dagger}$ &    --  & < $-0.95$  & $<-0.90$  & -- & $<-0.99$  & $<-0.92\, {\rm or}\, \gtrsim10^4$  \\
\hline
\vspace{0.10cm} 
 $\chi_{\rm min}^2$ & $-1421.826$ & $-1418.304$ & $-1418.220$ & $-1419.401$ & $-1413.071$ & $-1414.471$  \\ \vspace{0.10cm} 
$\Delta\chi_{\rm min}^2$ & $0$ & $-3.52$ & $-3.61$ & $0$ & $-6.33$ & $-4.93$ \\
  $\ln \mathcal{Z}$ & $-725.187$  & $-725.626$  & $-725.140$ &  $-728.869$  &  $-728.386$ & $-728.084$    \\
 $\Delta\ln\mathcal{Z}$ & $0$ & $0.44$ & $-0.05$ &  $0$ &  $-0.48$ & $-0.78$   \\ 
    \bottomrule
    \hline 
    \hline
  \end{tabular}
\end{table*}

\begin{figure}[t!]
\includegraphics[width=0.48\textwidth]{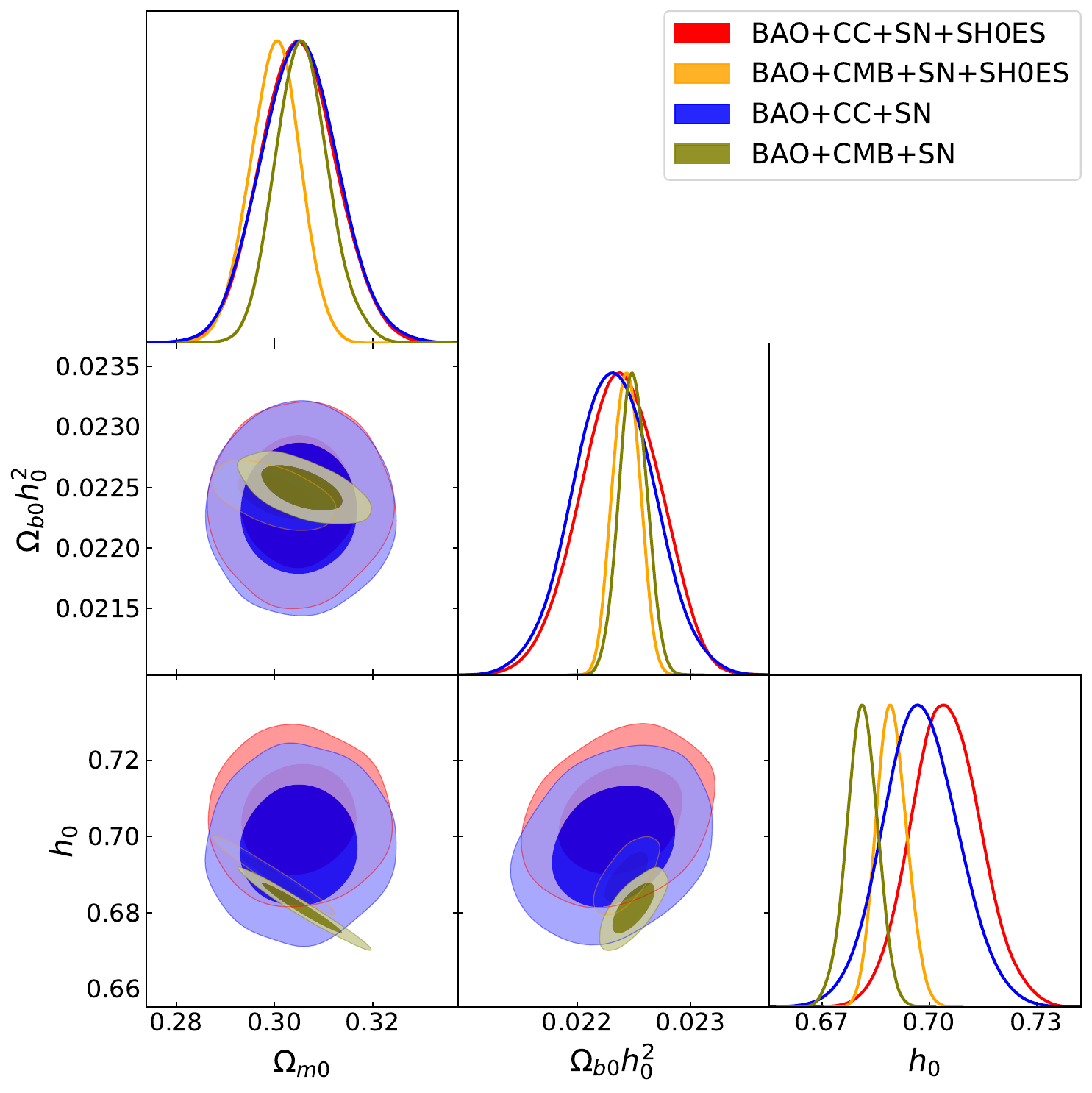}
  \caption{One- and two-dimensional ($68\%$  and  $95\%$  CLs)  marginalized posterior distributions for the free parameters of $\Lambda$CDM model using the combined datasets.}
  \label{fig:stdcontourlcdm}
\end{figure}

\begin{figure}[t!]
\includegraphics[width=0.48\textwidth]{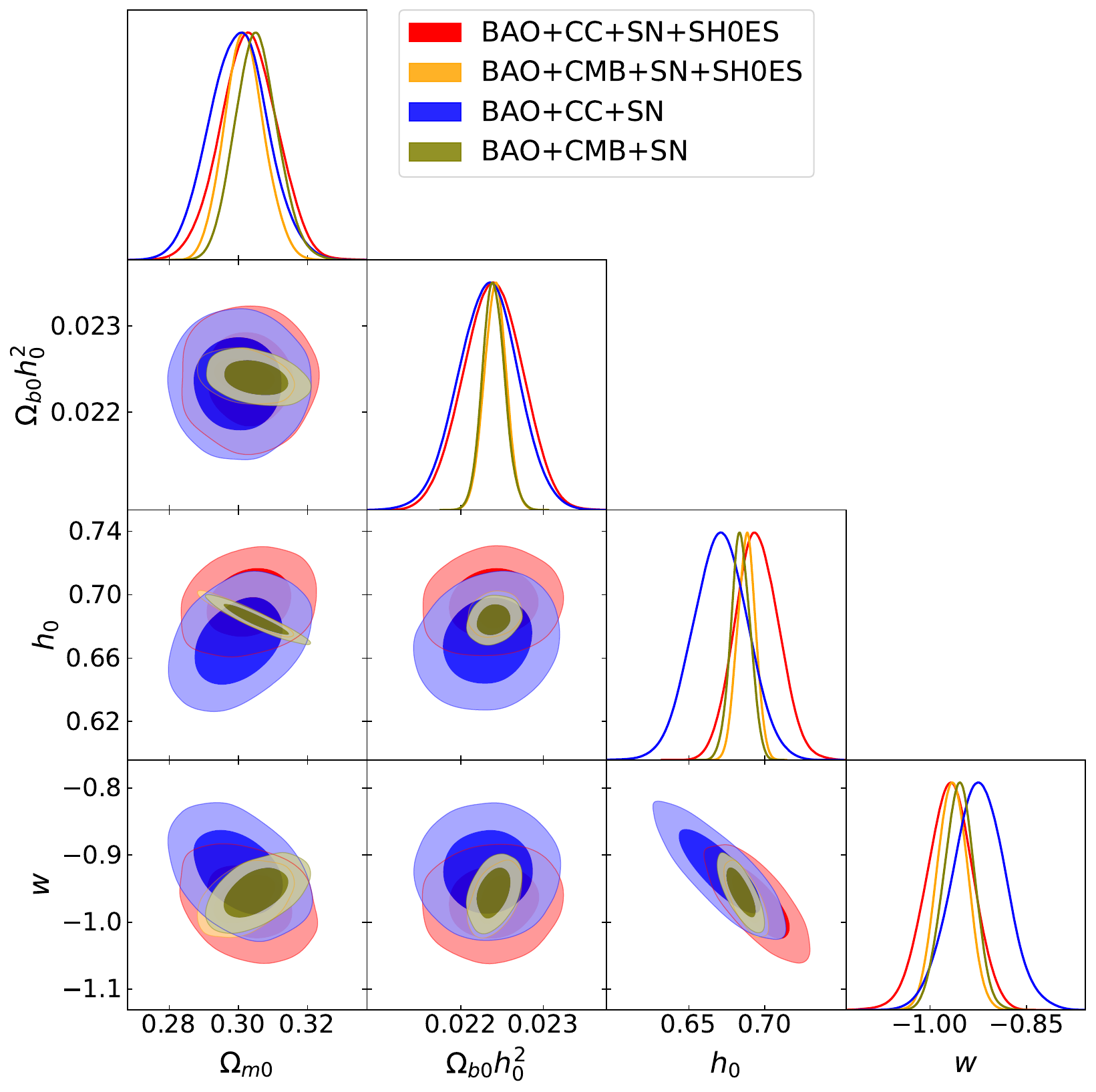}
  \caption{One- and two-dimensional ($68\%$  and  $95\%$  CLs)  marginalized posterior distributions for the free parameters of $w$CDM model using the combined datasets.}
  \label{fig:stdcontourwcdm}
\end{figure}

Figure~\ref{fig:stdcontourlcdm} illustrates the internal tension within the $\Lambda$CDM framework, where distinct observational datasets yield divergent constraints on $h_0$. Specifically, the Planck CMB measurements provide high-precision constraints that favor a lower $h_0$ value. In contrast, the inclusion of the SH0ES prior shifts the posterior toward higher values, underscoring the well-known discrepancy between early- and late-time probes. 
The contours in Fig.~\ref{fig:stdcontourwcdm} reveal a clear anti-correlation between $w$ and $h_0$, where shifting $w$ into the phantom regime ($w < -1$) drives the posterior toward higher $h_0$ values. This illustrates the well-known mechanism by which dark energy models with additional flexibility can alleviate the $H_0$ tension. 

\begin{figure}[t!]
\includegraphics[width=0.46\textwidth]{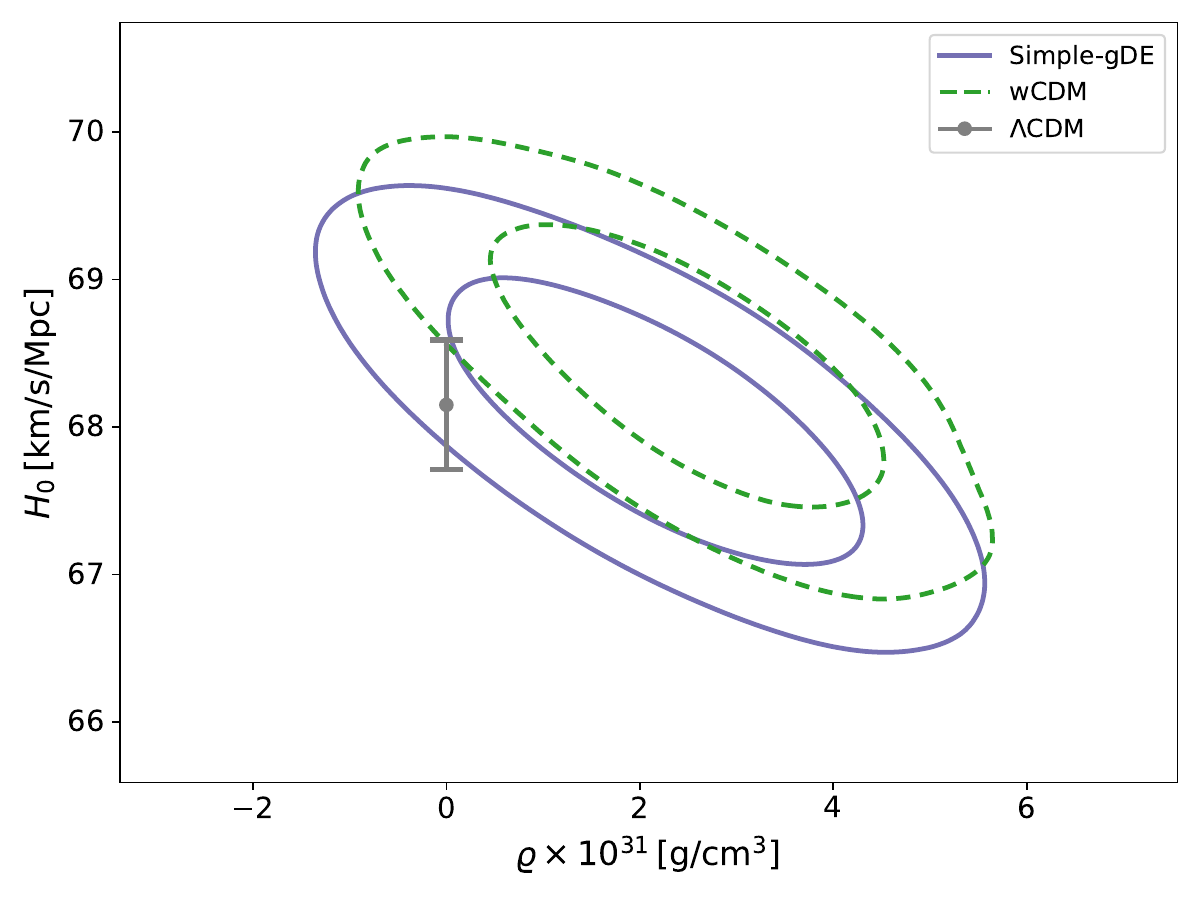}
  \caption{Joint constraints in the $(\varrho,H_0)$ plane for the Simple-gDE (solid blue) and $w$CDM (dashed green) models from BAO+CMB+SN dataset. Contours represent the $1\sigma$ confidence regions. The grey point with error bars indicates the reference $\Lambda$CDM value (fixed $\varrho=0$), $H_0=68.15\pm0.44\,\mathrm{km\,s^{-1}\,Mpc^{-1}}$.}
  \label{fig:nokH0varrhoPLK}
\end{figure}

Regarding the IMD, the combined BAO+CC+SN and BAO+CMB+SN datasets favor a slightly positive density (see Table~\ref{tab:noknoshoes}). However, the inclusion of the SH0ES prior (Table~\ref{tab:nokyesSHOES}) broadens the constraints to allow both positive and negative values, maintaining moderate statistical support for the $\Lambda$CDM baseline. Notably, the Planck CMB data permit more significant negative IMD values, a feature that typically acts to relax the $H_0$ tension. As illustrated in Fig.~\ref{fig:nokH0varrhoPLK}, while these negative densities provide some relief, they are not sufficient to fully reconcile the discrepancy with local measurements.

\begin{table*}[t!]
  \caption{Mean values and $68\%$ confidence-level uncertainties for the cosmological parameters in the $\Lambda$CDM, $w$CDM, and Simple gDE models constrained with the BAO+CC+SN+SH0ES and BAO+CMB+SN+SH0ES datasets. $\chi_{\rm min}^2=-2\ln{\mathcal{L}_{\rm max}}$ is used to compare best fit with respect to the $\Lambda$CDM model. The last rows contain the Bayesian evidence $\ln \mathcal{Z}$ and the relative Bayesian evidence with respect to the $\Lambda$CDM, i.e., $\Delta\ln \mathcal{Z}=\ln \mathcal{Z}_{{\Lambda}\rm CDM}-\ln \mathcal{Z}$.}
  \label{tab:nokyesSHOES}
   \begin{tabular}{lcccccc}
  	\hline
    \toprule
    \multicolumn{1}{l}{Dataset} & \multicolumn{3}{c}{\textbf{BAO+CC+SN+SH0ES}} & \multicolumn{3}{c}{\textbf{BAO+CMB+SN+SH0ES}} \\  \hline
& \textbf{$\Lambda$CDM} & \textbf{$w$CDM} & \textbf{Simple-gDE}
& \textbf{$\Lambda$CDM} & \textbf{$w$CDM} & \textbf{Simple-gDE} \\ 	\hline 
      \midrule
      \vspace{0.1cm}
$\Omega_{\rm m0}$ & $0.306\pm 0.009$ & $0.303\pm0.008$ & $0.303\pm0.008$ & $0.301\pm0.005 $ & $0.303\pm0.006$ & $0.307\pm0.018$ \\ \vspace{0.1cm}
$\Omega_{\rm b0}h_0^2$ &$0.0224\pm0.0004$& $0.0224\pm0.0004$ & $0.0224\pm0.0004$ & $ 0.0224\pm0.0001$ & $0.0224\pm 0.0001$ & $0.0225\pm0.0002$  \\ \vspace{0.1cm}
$h_0$  & $0.706\pm0.011$ & $0.695\pm0.014$  & $0.694\pm0.017$ & $0.689 \pm0.004 $  & $0.687\pm0.006$ & $0.681\pm0.021$  \vspace{0.1cm}\\
 $w_{\rm ci0}$  &  $-1$  & $-0.971\pm0.034$ &$-0.964\pm0.051$ &$-1$ &$-0.964\pm 0.021$  &  $-0.941\pm0.147$       \\ 
\hline
\vspace{0.1cm}
 $\varrho_{\rm ci0}\times 10^{31}$ $[\rm g\, cm^{-3}]$  &   0 & $1.66\pm2.16 $ & $ 2.28\pm2.95 $ & $0$ & $ 1.94\pm1.36 $  &  $ 3.19\pm5.08 $\\ \vspace{0.1cm}
$\Omega_{\rm ci0}$ & $0.650\pm0.008$  & $0.650\pm0.008$ & $0.651\pm0.009$ & $0.653\pm0.006$ & $0.651\pm0.007$ & $0.642\pm0.0228$  \\  \vspace{0.10cm} 
 $z_{\dagger}$ &    --  & < $-0.99\, {\rm or \,}\,\gtrsim 10^{28}$  & $<-0.98 \,\, {\rm or \,}\, \gtrsim 10^9$  & -- & $<-0.99$  & $<-0.99$   \\
\hline
\vspace{0.10cm} 
   $\chi_{\rm min}^2$ &  $-1424.234$ & $-1423.438$ & $-1423.408$ & $-1421.040$ & $-1417.758$ & $-1419.888$  \\ \vspace{0.10cm}
    $\Delta\chi_{\rm min}^2$ & $0$ & $-0.81$ & $-0.83$ & $0$ & $-3.28$ & $-1.15$ \\
  $\ln \mathcal{Z}$ & $-726.448$  & $-728.391$  & $-728.007$ &  $-729.716$      &     $-730.787$ & $-730.616$    \\
 $\Delta\ln \mathcal{Z}$ & 0 & $1.94$ & $1.56$ &  $0$      &     $1.07$ & $0.90$   \\ 
    \bottomrule
    \hline 
    \hline
  \end{tabular}
\end{table*}

\begin{figure}[t!]
\includegraphics[width=0.50\textwidth]{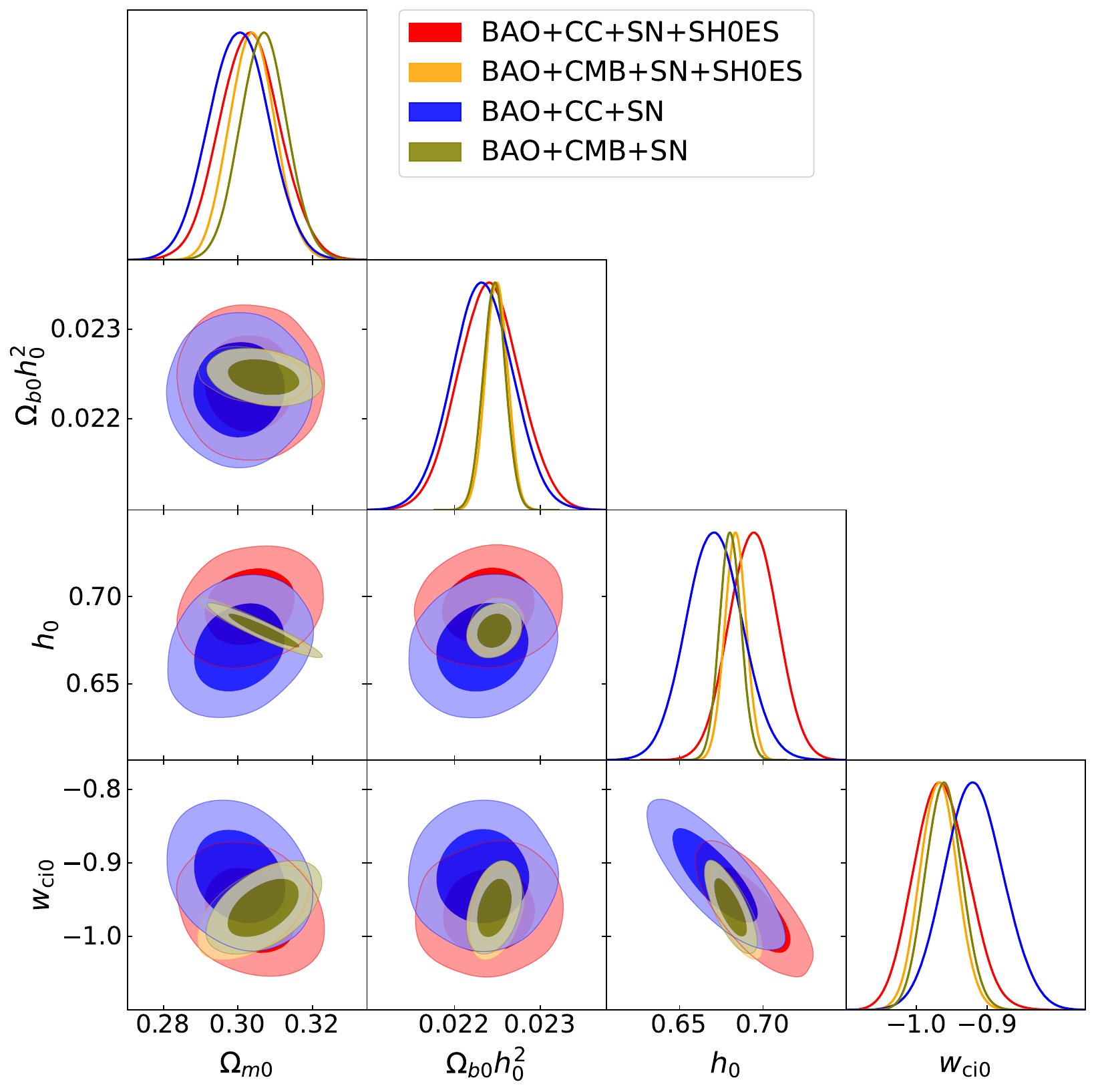}
  \caption{One- and two-dimensional ($68\%$  and  $95\%$  CLs)  marginalized posterior distributions for the free parameters of Simple-gDE model using the combined datasets.}
  \label{fig:stdcontour}
\end{figure}

Tables~\ref{tab:noknoshoes} and~\ref{tab:nokyesSHOES} demonstrate that when $w_{\rm ci0}$ (or $w$) is less than $-1$, the dark energy density $\rho$ necessarily changes sign in the past. However, because the observed deviation from the cosmological constant ($w=-1$) is minimal, Eq.~\eqref{eq:zpole} places the transition redshift at an extremely remote epoch ($z_{\rm ci\dagger} \sim 10^{28}$). Even when negative values for the inertial mass density are permitted at the $1\sigma$ level, their small magnitudes imply a transition redshift of $z_{\dagger} \gtrsim 10^{4}$. 
Such a transition would occur deep within the radiation-dominated era, significantly predating both recombination ($z \approx 1100$) and matter-radiation equality ($z \approx 3400$). Consequently, this occurs too early to meaningfully alter the late-time expansion history or drive a substantial increase in the inferred $H_0$. Conversely, if $w_{\rm ci0} > -1$ (a regime also mildly supported by the data) the IMD does not change sign in the past. In this case, the mathematical sign transition is projected into the far future (typically $z_{\rm ci\dagger} \approx -0.99$), which lacks physical relevance for our current cosmological horizon.

\begin{figure}[t!]
\includegraphics[width=0.46\textwidth]{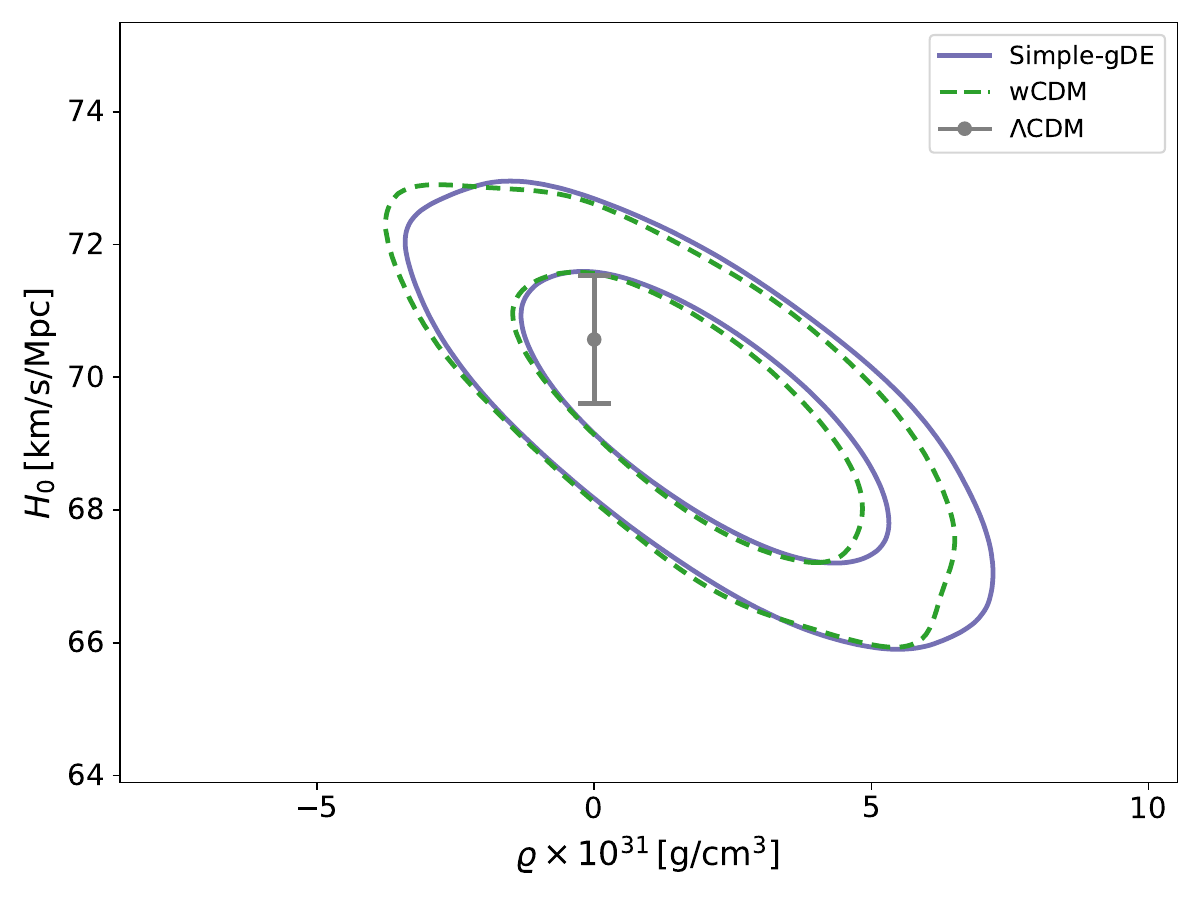}
  \caption{Joint constraints in the $(\varrho,H_0)$ plane for the Simple-gDE (solid blue) and $w$CDM (dashed green) models from BAO+CC+SN+SH0ES dataset. Contours represent the $1\sigma$ confidence regions. The grey point with error bars indicates the reference $\Lambda$CDM value (fixed $\varrho=0$), $H_0=70.57\pm0.96\,\mathrm{km\,s^{-1}\,Mpc^{-1}}$.}
  \label{fig:nokH0varrho}
\end{figure}

Based on the Bayesian model comparison, the evidence difference of $\Delta \ln \mathcal{Z} = -0.78$ is of order unity, suggesting no statistically significant preference between the model under consideration and the $\Lambda$CDM baseline. As shown in Table~\ref{tab:noknoshoes}, the inertial mass density for the combined BAO+CMB+SN dataset deviates from zero at nearly the $1\sigma$ confidence level in both the $w$CDM and Simple-gDE frameworks. While the Bayesian evidence slightly disfavors $w$CDM relative to $\Lambda$CDM, the difference is insufficient for decisive model selection.
For the BAO+CC+SN dataset, a dynamical equation-of-state model with $w_{\rm ci0} > -1$ (corresponding to a positive constant IMD) yields Bayesian evidence essentially equivalent to that of the $\Lambda$CDM model, which assumes a vanishing IMD. Similarly, the constant $w$CDM model with $w > -1$ provides a comparable fit to the data. However, as illustrated in Table~\ref{tab:nokyesSHOES} and the $H_0$ vs $\varrho$ 2D posterior in Fig.~\ref{fig:nokH0varrho}, the inclusion of the SH0ES prior causes the evidence to shift, with $\Lambda$CDM becoming moderately favored over both $w$CDM and Simple-gDE. 

\section{Negative energy density of DE via an interplay between non-null IMD and spatial curvature }
\label{sec:negativeDE}
The most significant finding of this work, particularly when allowing for non-zero spatial curvature, is that the best-fit trajectory suggests a \emph{Null Energy Condition (NEC) crossing} occurring simultaneously with a \emph{sign change in the dark energy density} in the past. This dynamical evolution is of particular interest as it aligns with recent indications from DESI measurements \cite{DESI:2024mwx,DESI:2025zgx}. In the existing literature, similar phenomenology can occur within the CPL parametrization for the parameter space $w_0>-1$ (positive inertial mass density today) and $w_a<0$, which yields a negative inertial mass density in the past. Remarkably, however, our model further predicts that the dark energy density $\rho_{\rm DE}$ itself evolves from negative to positive values. This specific transition is absent in the standard CPL framework but remains a defining characteristic of $\Lambda_{\rm s}$CDM-type models~\cite{Alexandre:2023nmh,Akarsu:2025gwi,Akarsu:2025dmj,Akarsu:2024qsi,Akarsu:2024eoo,Akarsu:2024nas,Souza:2024qwd,DiGennaro:2022ykp,Nyergesy:2025lyi,Anchordoqui:2023woo,Anchordoqui:2024gfa,Anchordoqui:2024dqc,Soriano:2025gxd}

\begin{table*}[ht!]
  \caption{Mean values and $68\%$ confidence-level uncertainties for the cosmological parameters in the $o\Lambda$CDM, $ow$CDM, and $o$Simple gDE models constrained with the BAO+CC+SN and BAO+CMB+SN datasets. $\chi_{\rm min}^2=-2\ln{\mathcal{L}_{\rm max}}$ is used to compare best fit with respect to the $\Lambda$CDM model. The last rows contain the Bayesian evidence $\ln \mathcal{Z}$ and the relative Bayesian evidence with respect to the $\Lambda$CDM, i.e., $\Delta\ln \mathcal{Z}=\ln \mathcal{Z}_{{\Lambda}\rm CDM}-\ln \mathcal{Z}$. }
  \label{tab:yesknoSh0es}
  \begin{tabular}{lcccccc}
    \hline
    \toprule
    \multicolumn{1}{l}{Dataset} & \multicolumn{3}{c}{\textbf{BAO+CC+SN}} & \multicolumn{3}{c}{\textbf{BAO+CMB+SN}} \\
    \hline
    & \textbf{{o$\Lambda$CDM}} &  $\quad$ \textbf{o$w$CDM} & $\quad$\textbf{osimple-gDE} & $\quad$ $\quad$\textbf{o{$\Lambda$CDM}} &$\quad$ \textbf{o$w$CDM} &$\quad$ \textbf{osimple-gDE}  \\
    \hline 
    \midrule
    $\Omega_{\rm m0}$ & $0.290 \pm 0.011$ & $0.302 \pm 0.011$ & $0.301 \pm 0.009$ & $0.301 \pm 0.006$ & $0.304 \pm 0.007$ & $0.305 \pm 0.009$ \\
    $\Omega_{\rm b0}h_0^2$ & $0.0223 \pm 0.0004$ & $0.0223 \pm 0.0004$ & $0.0223 \pm 0.0004$ & $0.0223 \pm 0.0001$ & $0.0224 \pm 0.0001$ & $0.0224 \pm 0.0002$ \\
    $h_0$ & $0.672 \pm 0.017$ & $0.671 \pm 0.019$ & $0.671 \pm 0.018$ & $0.692 \pm 0.007$ & $0.686 \pm 0.008$ & $0.684 \pm 0.011$ \\
    $w_{\rm kci0}$ & $-1$ & $-0.911 \pm 0.047$ & $-0.914 \pm 0.051$ & $-1$ & $-0.956 \pm 0.025$ & $-0.947 \pm 0.060$ \\
    $\Omega_{k0}$ & $0.069 \pm 0.035$ & $-0.017 \pm 0.031$ & $-0.002 \pm 0.018$ & $0.002 \pm 0.002$ & $0.003 \pm 0.002$ & $0.003 \pm 0.002$ \\
    \hline
    $\varrho_{\rm kci}\times 10^{31}$ $[\rm g\, cm^{-3}]$  & 0 & $4.08\pm2.79$ & $4.11 \pm 3.19$ & 0 & $2.51\pm1.43$  & $3.31 \pm 3.11$ \\
    $\Omega_{\rm ci0}(\Omega_{\Lambda})\,(\Omega_{w})$ & $0.612\pm0.027$ &  $0.649\pm0.033$ & $0.643\pm 0.039$ & $0.652\pm0.006$  & $0.645\pm 0.007$ & $0.643\pm0.013$ \\
   $\Omega_{\rm kci0}(\Omega_{o\Lambda})\,(\Omega_{ow})$ & $0.657\pm0.011$ & $0.651\pm0.010$ & $0.651\pm0.011$ & $0.653\pm0.007$  & $0.648\pm0.008$ & $0.646\pm0.012$ \\
    $z_{{\rm kci\dagger}}$ ($68\%$) & $7.89^{+7.70}_{-2.82}$ & $30.28^{+28.77}_{-11.45}$  &$6.05^{+8.70}_{-3.19}$ & $27.70^{+26.50}_{-9.86}$ & --  & --\\
    \hline
    $\chi_{\rm min}^2$ & $-1419.980$ & $-1418.288$ & $-1418.194$ & $-1415.972$ & $-1412.902$ & $-1412.888$ \\
    $\Delta \chi_{\rm min}^2$ & $0$ & $-1.69$ & $-1.79$ & $0$ & $-3.07$ & $-3.08$ \\
    $\ln \mathcal{Z}$ & $-726.695$ & $-727.697$ & $-727.289$ & $-732.443$ & $-733.567$ & $-732.467$ \\
    $\Delta\ln \mathcal{Z}$ & $0$ & $1.01$ & $0.59$ & $0$ & $1.12$ & $0.02$ \\
    \bottomrule
    \hline
    \hline
  \end{tabular}
\end{table*}

\begin{figure}[t!]
\includegraphics[width=0.5\textwidth]{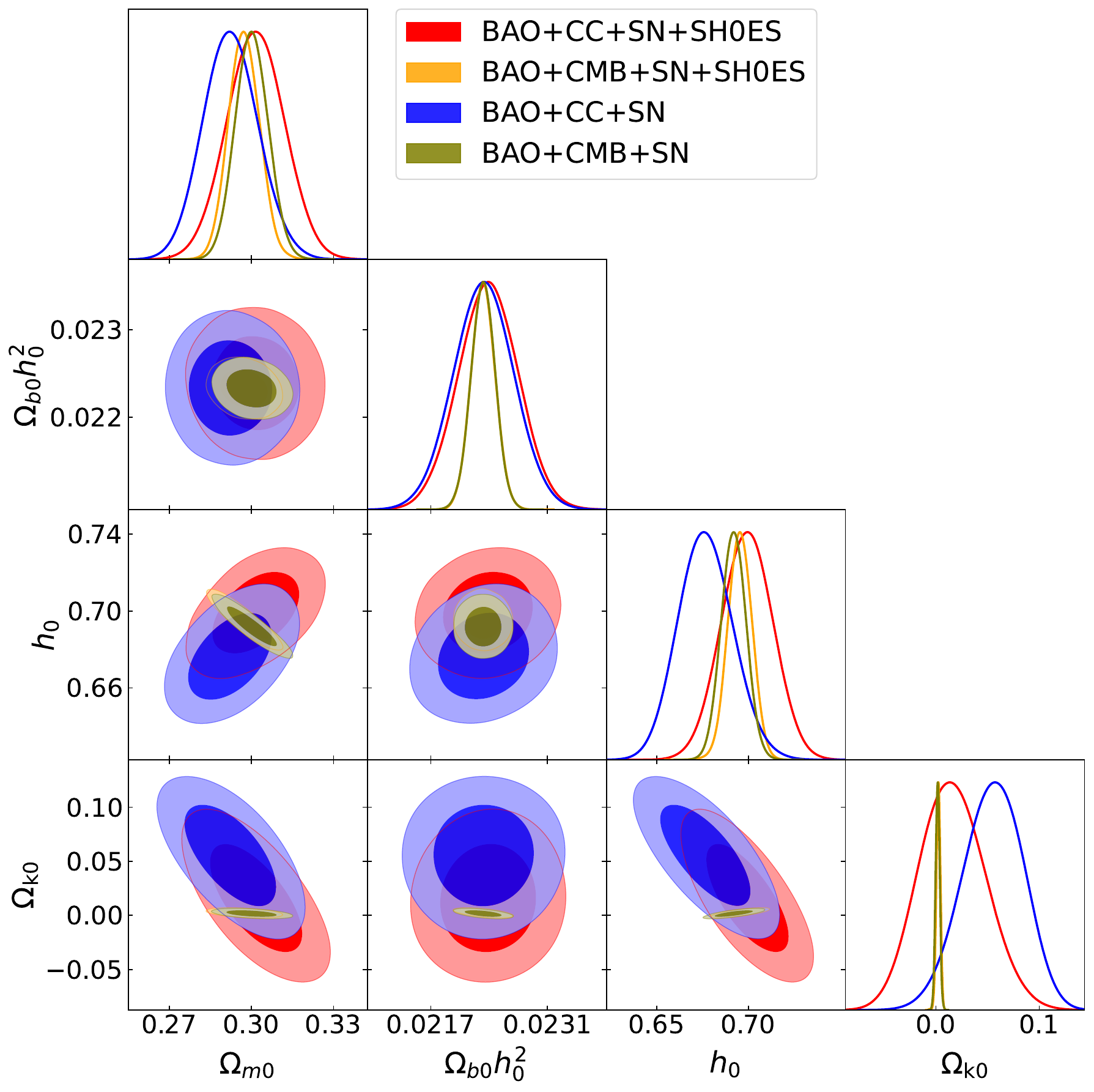}
  \caption{One- and two-dimensional ($68\%$  and  $95\%$  CLs)  marginalized posterior distributions for the free parameters of $o\Lambda$CDM model using the combined datasets.}
  \label{fig:olcdm}
\end{figure} 

\begin{figure}[t!]
\includegraphics[width=0.5\textwidth]{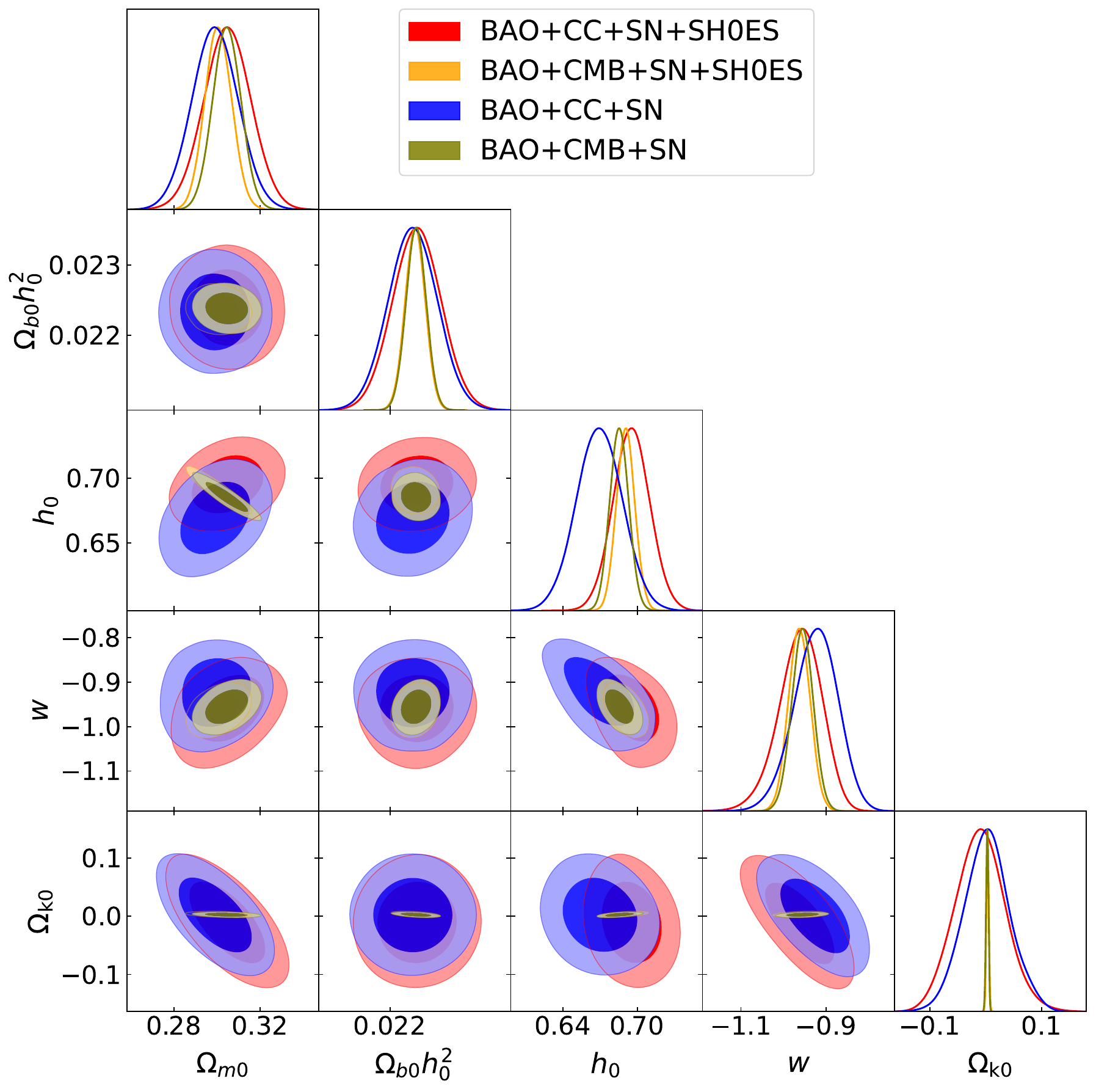}
  \caption{One- and two-dimensional ($68\%$  and  $95\%$  CLs)  marginalized posterior distributions for the free parameters of $ow$CDM model using the combined datasets. }
  \label{fig:owcdm}
\end{figure}

Theoretical considerations indicate that if $1+w_{\rm DE0} > 0$ (provided that $\rho_{\rm DE}$ has already transitioned to positive values) a negative spatial curvature parameter ($\Omega_{k0} < 0$) is required. This implies that a model characterized by a \emph{constant positive inertial mass density} can, in principle, manifest these two dynamical features only within a \emph{spatially closed} Universe.
As discussed in Sec.~\ref{sec:curvatureDE}, a sign transition in the dark energy density is theoretically viable in the $o\Lambda$CDM, $ow$CDM, and $o$Simple-gDE frameworks, facilitated by the inclusion of spatial curvature. Conversely, a sign transition in the inertial mass density itself is only possible within the $ow$CDM and $o$Simple-gDE models. In this section, we examine the observational implications for these models to identify which configurations are statistically favored when spatial curvature is treated as a free parameter.

Table~\ref{tab:yesknoSh0es} provides a comparison of cosmological constraints derived from the BAO+CC+SN and BAO+CMB+SN datasets across the $o\Lambda$CDM, $ow$CDM, and $o$Simple-gDE frameworks. For the BAO+CC+SN combination, the Bayesian evidence, $\Delta \ln \mathcal{Z}=1.01$ for $ow$CDM and $\Delta \ln \mathcal{Z}=0.59$ for $o$Simple-gDE, indicates that the $o\Lambda$CDM model remains favored according to the Jeffreys scale. Specifically, the value $\Delta \ln \mathcal{Z}=0.59$ suggests that the $o$Simple-gDE extension is statistically indistinguishable from the $o\Lambda$CDM baseline. 
The inclusion of spatial curvature as a free parameter reveals a mild, $\sim2\sigma$ preference for an open universe ($\Omega_{k0}>0$) within the $o\Lambda$CDM framework for the BAO+CC+SN dataset. Conversely, in the $ow$CDM and $o$Simple-gDE models, the equation-of-state parameters are constrained to values slightly exceeding $-1$, implying a positive inertial mass density today. In these cases, the spatial curvature constraints broaden, remaining consistent with both open and closed geometries within the $1\sigma$ confidence level. 

The inclusion of Planck CMB data significantly tightens the resulting parameter constraints. Within the $o\Lambda$CDM framework, the Hubble parameter is constrained to $h_0 = 0.692 \pm 0.007$, while the curvature parameter is driven toward spatial flatness, with $\Omega_{k0} \in [0, 0.04]$ taking the $1\sigma$ constrain. Notably, the dynamical models yield slightly lower values for the Hubble parameter: $h_0 = 0.686 \pm 0.008$ for $ow$CDM and $h_0 = 0.684 \pm 0.011$ for $o$Simple-gDE. 
This shift occurs because the additional degrees of freedom in the $ow$CDM and $o$Simple-gDE models allow for open-universe solutions, which are physically correlated with lower $H_0$ values. From a model-selection perspective, the Bayesian evidence differences relative to $o\Lambda$CDM are $\Delta \ln \mathcal{Z} = 1.12$ for $ow$CDM and $\Delta \ln \mathcal{Z} = 0.02$ for $o$Simple-gDE. These results indicate that the data remain fully compatible with the $o\Lambda$CDM scenario; there is only weak evidence against $ow$CDM, and $o$Simple-gDE remains statistically indistinguishable from the baseline.

The inclusion of the SH0ES prior (Table~\ref{tab:yeskyesSh0es}) introduces notable shifts in the inferred constraints, which may initially seem to diverge from theoretical expectations. Within the $o\Lambda$CDM framework, Planck CMB measurements continue to drive the parameters toward spatial flatness, with only a marginal preference for positive spatial curvature ($\Omega_{k0} = 0.021 \pm 0.036$) at the $1\sigma$ level (see Fig.~\ref{fig:olcdm}). Due to the anti-correlation between $\Omega_{k0}$ and $h_0$—primarily driven by the cosmic chronometers and Pantheon+ datasets—this slight curvature preference results in lower $h_0$ values. Consequently, this configuration does not support a sign transition in the dark energy density; furthermore, as previously established, the $o\Lambda$CDM model precludes a sign transition in the inertial mass density by construction.
The $ow$CDM model allows for a density sign transition at $z_{\dagger} \sim 4.5$ (Fig.~\ref{fig:owcdm} and Table~\ref{tab:yeskyesSh0es}). However, despite this additional dynamical flexibility, the Bayesian evidence yields $\Delta \ln \mathcal{Z} = 1.65$ relative to $o\Lambda$CDM. According to the Jeffreys scale, this indicates moderate statistical support for the simpler $o\Lambda$CDM model over the $ow$CDM extension, suggesting that the current data do not yet require the complexity of a sign-changing dark energy density.

\begin{figure}[t!]
\includegraphics[width=0.5\textwidth]{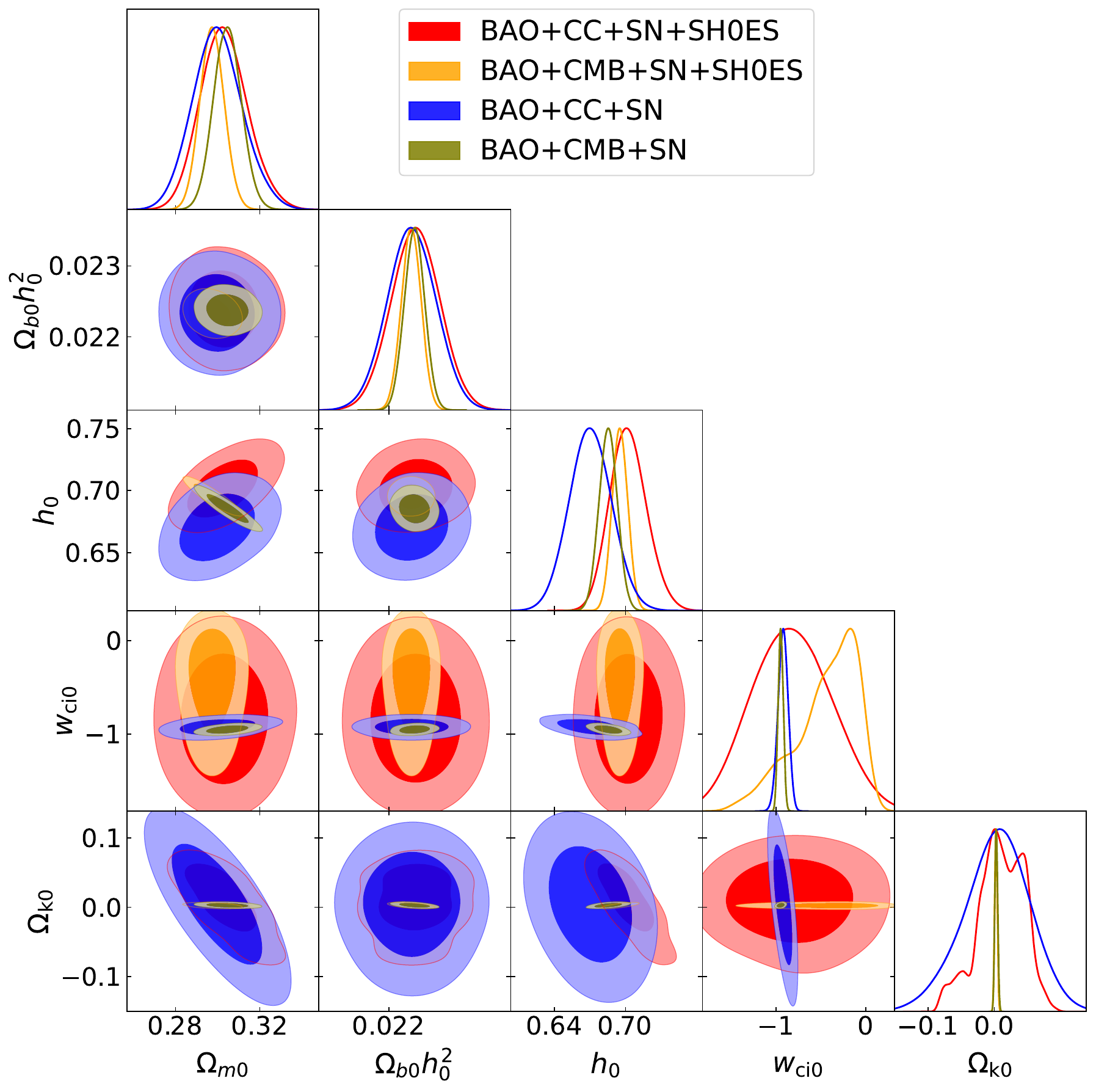}
  \caption{One- and two-dimensional ($68\%$  and  $95\%$  CLs)  marginalized posterior distributions for the free parameters of $o$SimpleGDE model using the combined datasets.}
  \label{fig:oSgDE}
\end{figure}

In contrast, as illustrated in Fig.~\ref{fig:oSgDE}, the SH0ES prior drives the posterior toward a significantly higher value of $H_0 = 71.72 \pm 1.35\,\text{km}\,\text{s}^{-1}\,\text{Mpc}^{-1}$ for the BAO+CC+SN+SH0ES combination. When Planck data are integrated, the inferred value remains elevated at $H_0 = 69.78 \pm 1.51\,\text{km}\,\text{s}^{-1}\,\text{Mpc}^{-1}$. Notably, for the BAO+CC+SN+SH0ES dataset, the Bayesian evidence difference of $\Delta \ln \mathcal{Z} = -3.63$ constitutes strong evidence in favor of the $o$Simple-gDE model over the $o\Lambda$CDM baseline. 
A comparison of the contours in Fig.~\ref{fig:owcdm} and Fig.~\ref{fig:oSgDE} reveals that while both models exhibit a similar degeneracy between the expansion rate and the dark energy parameters ($w$ or $w_{\rm ci0}$), the underlying physical mechanisms remain distinct. In the $w$CDM model, the $H_0$ shift is driven purely by a constant equation-of-state parameter $w$. Conversely, in the $o$Simple-gDE framework, the expansion history is governed by a constant IMD. This allows the data to explore both positive and negative IMD values, enabling a sign transition in the effective sum $\rho_{\rm DE} + p_{\rm DE}$, a dynamical feature fundamentally absent in the standard $w$CDM parametrization.

\subsubsection{Departure from Gaussian Posterior Distribution}

\begin{figure}[t!]
\includegraphics[width=0.47\textwidth]{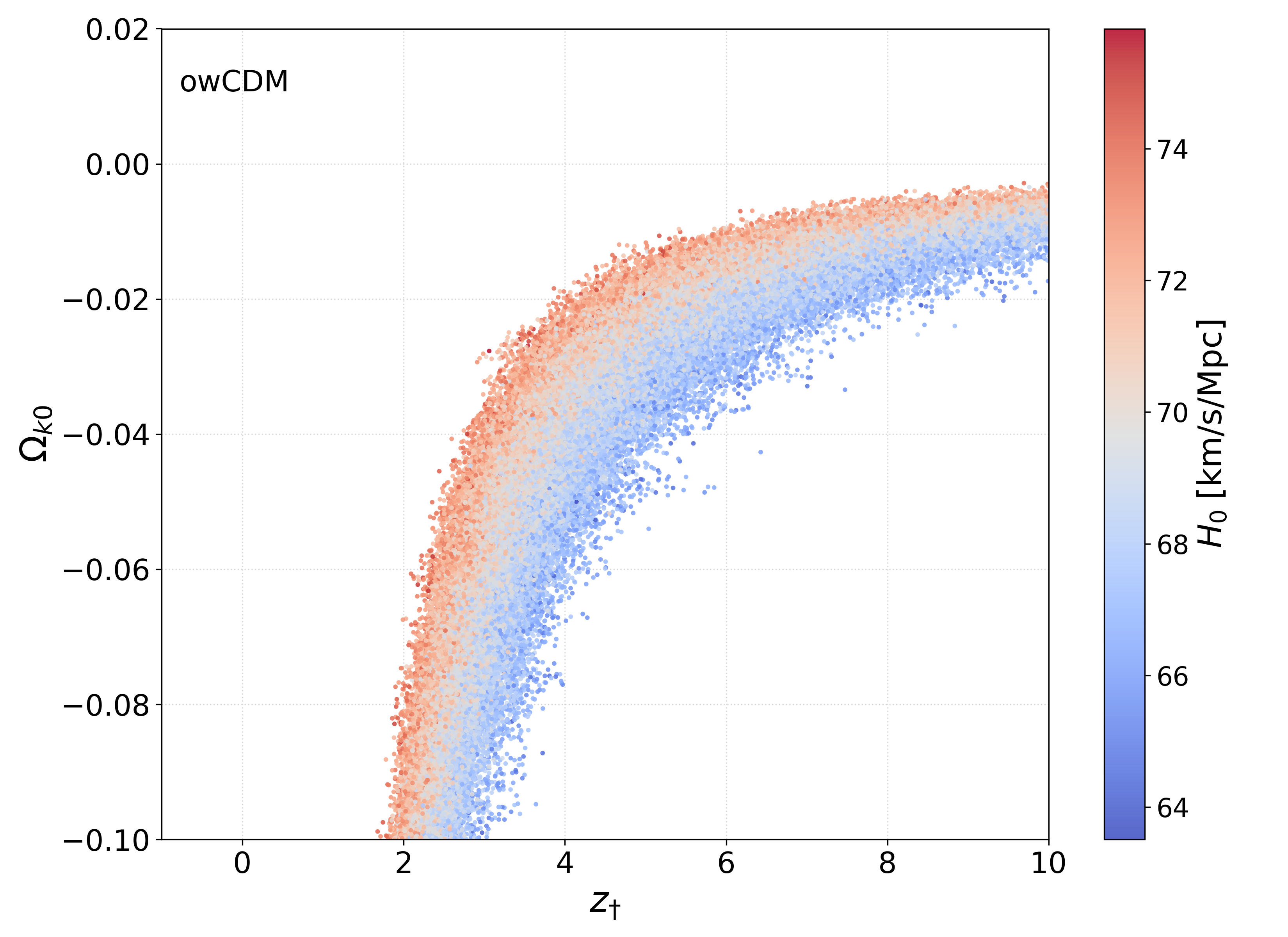}
  \caption{Scatter plot of the transition redshift $z_{\dagger}$ versus $\Omega_{k0}$ for the $ow$CDM model given in Eq.\eqref{eq:owdagger}. 
Points are colored according to $H_0$ for the BAO+CC+SN+SH0ES data set.}
  \label{fig:Okh0owcdm}
\end{figure}

\begin{table*}[ht!]
  \caption{Mean values and $68\%$ confidence-level uncertainties for the cosmological parameters in the $o\Lambda$CDM, $ow$CDM, and $o$Simple gDE models constrained with the BAO+CC+SN+SH0ES and BAO+CMB+SN+SH0ES datasets. $\chi_{\rm min}^2=-2\ln{\mathcal{L}_{\rm max}}$ is used to compare best fit with respect to the $\Lambda$CDM model. The last rows contain the Bayesian evidence $\ln \mathcal{Z}$ and the relative Bayesian evidence with respect to the $\Lambda$CDM, i.e., $\Delta\ln \mathcal{Z}=\ln \mathcal{Z}_{{\Lambda}\rm CDM}-\ln \mathcal{Z}$.}
  \label{tab:yeskyesSh0es}
  \begin{tabular}{lcccccc}
    \hline
    \toprule
    \multicolumn{1}{l}{Dataset} & \multicolumn{3}{c}{\textbf{BAO+CC+SN+SH0ES}} & \multicolumn{3}{c}{\textbf{BAO+CMB+SN+SH0ES}} \\
    \hline
    & \textbf{{o$\Lambda$CDM}} &  $\quad$ \textbf{o$w$CDM} & $\quad$\textbf{osimple-gDE} & $\quad$ $\quad$\textbf{o{$\Lambda$CDM}} &$\quad$ \textbf{o$w$CDM} &$\quad$ \textbf{osimple-gDE}  \\
    \hline 
    \midrule
    $\Omega_{\rm m0}$ & $0.301\pm0.011$ & $0.309\pm0.011$ & $0.313\pm 0.011$ & $ 0.297\pm0.006 $ & $0.301\pm0.006$ & $0.298\pm0.006$ \\
    $\Omega_{\rm b0}h_0^2$ & $0.0224\pm0.0004$ & $0.0224\pm0.0004$ & $0.0224\pm0.0004$ & $0.0223\pm0.0001$ & $0.0224\pm0.0001$ & $0.0223\pm0.0001$ \\
    $h_0$ & $0.698\pm0.015$ & $0.697\pm0.015$ & $0.717\pm0.014$ & $0.696\pm0.007$ & $0.691\pm0.007$ & $0.695\pm 0.006$ \\
    $w_{\rm kci0}$ & $-1$ & $-0.941\pm0.047$ & $-0.536\pm 0.228$ & $-1$ & $-0.963\pm0.025$ & $-0.641\pm 0.257$ \\
    $\Omega_{k0}$ & $0.021\pm0.036$ & $ -0.036\pm0.041$ & $-0.035\pm0.026$ & $0.002\pm0.002$ & $0.003\pm0.002$ & $0.002\pm0.002$ \\
    \hline
    $\varrho_{\rm kci}\times10^{31}$ $[\rm g\, cm^{-3}]$  & $0$ & $2.296\pm3.091$ & $18.6^{+27.1}_{-29.9}$ & $0$ & $2.122\pm1.471$  &  $43.6^{+11.9}_{-26.1}$    \\
    $\Omega_{\rm ci0}$ & $0.639\pm0.025$ &  $0.658\pm0.037$  & $0.641\pm0.027$ & $0.654\pm0.006$  & $0.649\pm0.007$ & $0.654\pm0.006$ \\ 
    $\Omega_{\rm kci0}$ & $0.652\pm0.011$ & $0.649\pm0.011$  & $0.652\pm0.010$ & $0.656\pm0.007$  & $0.652\pm0.007$ & $0.656\pm0.007$ \\
    $z_{{\rm kci\dagger}}\,(>0)(68\%)$ & $6.21^{+5.41}_{-2.27}$ & $4.49^{+4.25}_{-1.63}$ & $1.51^{+0.68}_{-0.34}$ $\&$ $13.1^{+14.5}_{-6.6}$  & $30.28^{+28.77}_{-11.45}$ & -- & $35.1^{+37.1}_{-13.0}$ \\
     \hline
    $\chi_{\rm min}^2$ & $-1423.972$ & $-1423.371$ & $-1417.546$ & $-1419.371$ & $-1417.201$ & $-1419.361$ \\
    $\Delta \chi_{\rm min}^2$ & $0$ & $-0.60$ & $-6.43$ & $0$ & $-2.17$ & $-0.01$ \\
    $\ln \mathcal{Z}$ & $-728.857$ & $-730.511$ & $ -725.222$ & $-734.211$ & $-735.773$ & $-733.815$ \\
    $\Delta\ln \mathcal{Z}$ & $0$ & $1.65$ & $-3.63$ & 0 & $1.56$ & $-0.395$ \\
    \bottomrule
    \hline
    \hline
  \end{tabular}
\end{table*}

\begin{table*}[t]
\centering
\caption{Constraints on the transition redshift $z_{\dagger}$ obtained from different dataset combinations. 
We report both percentile-based Bayesian credible intervals and Gaussian approximations derived from the MCMC posterior distributions.}
\begin{tabular}{lcccc}
\hline
Dataset & Percentile (68\% CL) & Percentile (95\% CL) & Gaussian ($1\sigma$) & Gaussian ($2\sigma$) \\
\hline
BAO+CC+SN ($o\Lambda$CDM) 
& $7.89^{+7.70}_{-2.82}$ 
& $7.89^{+33.18}_{-4.01}$ 
& $11.64 \pm 17.72$ 
& $11.64 \pm 35.44$ \\

BAO+CC+SN+SH0ES ($o\Lambda$CDM) 
& $6.21^{+5.41}_{-2.27}$ 
& $6.21^{+23.77}_{-3.33}$ 
& $8.79 \pm 13.68$ 
& $8.79 \pm 27.36$ \\

BAO+CMB+SN ($o\Lambda$CDM) 
& $27.70^{+26.50}_{-9.86}$ 
& $27.70^{+115.63}_{-14.39}$ 
& $40.55 \pm 58.94$ 
& $40.55 \pm 117.88$ \\

BAO+CC+SN ($ow$CDM) 
& $30.28^{+28.77}_{-11.45}$ 
& $30.28^{+126.46}_{-17.25}$ 
& $43.45 \pm 55.18$ 
& $43.45 \pm 110.35$ \\

BAO+CC+SN+SH0ES ($o$Simple-gDE $W_0$) 
& $1.51^{+0.68}_{-0.34}$ 
& $1.51^{+2.22}_{-0.49}$ 
& $1.76 \pm 2.38$ 
& $1.76 \pm 4.77$ \\

BAO+CC+SN+SH0ES ($o$Simple-gDE $W_{-1}$) 
& $13.1^{+14.5}_{-6.6}$ 
& $13.1^{+61.4}_{-9.7}$ 
& $20.2 \pm 50.8$ 
& $20.2 \pm 101.6$ \\

BAO+CMB+SN+SH0ES ($o$Simple-gDE $W_{-1}$) 
& $35.1^{+37.1}_{-13.0}$ 
& $35.1^{+163.2}_{-18.6}$ 
& $53.7 \pm 89.6$ 
& $53.7 \pm 179.2$ \\

BAO+CMB+SN ($o$Simple-gDE $W_{-1}$) 
& $6.05^{+8.70}_{-3.19}$ 
& $6.05^{+34.55}_{-3.84}$ 
& $10.16 \pm 23.20$ 
& $10.16 \pm 46.40$ \\

\hline
\end{tabular}
\label{tab:zdagger_all}
\end{table*}

\begin{figure}[t!]
\includegraphics[width=0.5\textwidth]{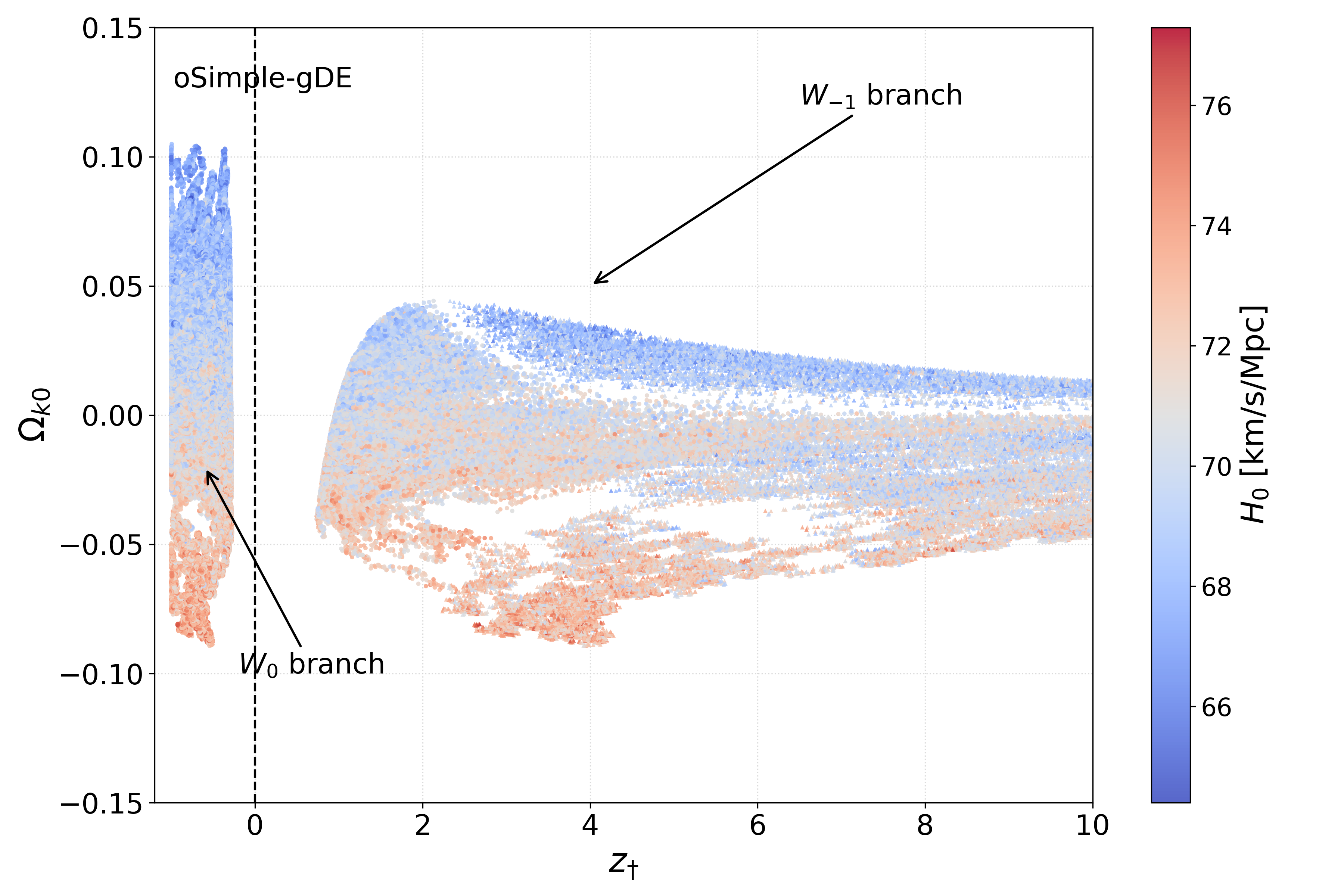}
  \caption{Scatter plot of the transition redshift $z_{\dagger}$ versus $\Omega_{k0}$ for the $o$Simple-gDE model given in \eqref{eq:tildezpole}. 
Points are colored according to $H_0$. 
The vertical dashed line marks $z_{\dagger}=0$. 
Future transitions ($z_{\dagger}<0$) arise only from the principal Lambert branch $W_0$, whereas the $W_{-1}$ branch produces only $z_{\dagger}\ge0$ solutions, namely sign transition in the past near $z\sim2$ for the BAO+CC+SN+SH0ES data set. }
  \label{fig:Okh0}
\end{figure}

The transition redshift $z_{\dagger}$ (at which the effective dark energy density changes sign) is computed as a derived parameter from the MCMC chains. For $o$Simple-gDE model, as seen from  Figure \ref{fig:Okh0}, the calculation involves evaluating the two real branches of the Lambert-$W$ function ($W_0$ and $W_{-1}$). 
Only physically acceptable solutions are retained, requiring the existence of real branches ($x\in[-1/e,0)$) together with the condition $z_{\dagger}>0$. The constraints are obtained directly from the posterior samples without rerunning the MCMC chains. 
Instead, physical cuts are applied to the existing samples, and summary statistics are recomputed using the resulting subsample.

Uncertainties are estimated using two approaches: (i) the Gaussian approximation, expressed as mean $\pm$ standard deviation, and (ii) percentile-based Bayesian credible intervals (68\% and 95\%). 
Due to the asymmetric and long-tailed nature of the distribution, percentile-based intervals provide a more reliable statistical description. Therefore, the final results for $z_{\dagger}$ given in Tables \ref{tab:yesknoSh0es} and \ref{tab:yeskyesSh0es} are reported using a percentile-based masking procedure. 

For completeness, in Table~\ref{tab:zdagger_all} we report both percentile-based Bayesian credible intervals and Gaussian approximations for $z_{\dagger}$. However, as anticipated from the strong dependence of the transition redshift on the curvature parameter near spatial flatness, the posterior distributions exhibit significant skewness and extended high-redshift tails. In this regime, the Gaussian (mean $\pm \sigma$) approximation—which implicitly assumes a symmetric and unimodal distribution—is statistically inadequate. This non-Gaussianity is driven by the non-linear Lambert-$W$ dependence (specifically the $W_{-1}$ branch) and the imposition of physical cuts. Consequently, while the Gaussian estimate provides a misleading description of the uncertainty, the non-parametric percentile-based intervals capture the asymmetric structure of the posterior more faithfully, particularly for solutions approaching the $\Lambda$CDM limit where $z_{\dagger}$ formally diverges.

Therefore, throughout this work we adopt percentile-based credible intervals as our primary summary of the constraints. 
Figures~\ref{fig:Okh0owcdm} and \ref{fig:Okh0} explicitly indicate that a slight negative contribution to the energy density from closed spatial curvature enables a sign transition in the effective dark energy density in the past for both the $ow$CDM and $o$Simple-gDE models. The posterior distributions for the transition redshift parameter $z_{\dagger}$ are markedly non-Gaussian. This behavior is not immediately apparent from the tabulated $z_{\dagger}$ values in Tables~\ref{tab:yesknoSh0es} and \ref{tab:yeskyesSh0es}.
The long tail of the posterior distribution toward large values effectively corresponds to solutions approaching the $\Lambda$CDM limit ($z_{\dagger} \to \infty$). This leads to an asymmetric distribution with an unbounded upper bound, making a simple Gaussian approximation based on the mean and standard deviation statistically inadequate. Such behavior is a generic feature of models allowing a transition from positive to negative energy density, as seen in $\Lambda_{\rm s}$CDM-type scenarios~\cite{Akarsu:2021fol,Akarsu:2022typ,Akarsu:2023mfb,Escamilla:2025imi}.

The most promising result, $o$Simple-gDE model in  $W_0$ region with the contraint 
 $z_{\dagger}=1.51^{+0.68}_{-0.34}$ 
for the BAO+CC+SN+SH0ES dataset. As stated above, the Bayesian evidence difference $\Delta \ln \mathcal{Z} = -3.63$ provides strong support for the $o$Simple-gDE model relative to $o\Lambda$CDM. Conversely, the comparison with the $ow$CDM model yields $\Delta \ln \mathcal{Z} = 1.65$, indicating moderate evidence in favor of $o\Lambda$CDM. 
Overall, these results highlight that the data significantly prefer the $o$Simple-gDE scenario over $o\Lambda$CDM, while the standard $o\Lambda$CDM model remains moderately favored over $ow$CDM. These results also demonstrate that spatial curvature can substantially modify the inferred dark-energy phenomenology. Even a small negative curvature contribution effectively shifts the dark-energy sector, allowing a transition in the effective energy density that is absent in the flat $\Lambda$CDM limit. 
This illustrates that relaxing the assumption of spatial flatness can qualitatively change the cosmological interpretation of current data and the relative performance of dark-energy models.

\begin{table*}[t!]
\centering
\caption{Constraints on the inertial mass density parameter $\varrho$ in $o$Simple-gDE model obtained from the MCMC posterior distributions for different dataset combinations. Both percentile-based credible intervals and Gaussian approximations are reported.}
\begin{tabular}{lcccc}
\hline
Dataset & $\varrho$  Percentile (68\% CL) & $\varrho$  Percentile (95\% CL) & $\varrho$  Gaussian ($1\sigma$) & $\varrho$ Gaussian ($2\sigma$) \\
\hline

BAO+SN+CMB+SH0ES 
& $(4.36^{+1.19}_{-2.61})\times10^{-30}$ 
& $(4.36^{+1.57}_{-5.51})\times10^{-30}$ 
& $(3.75 \pm 1.97)\times10^{-30}$ 
& $(3.75 \pm 3.94)\times10^{-30}$ \\

BAO+SN+CC+SH0ES 
& $(1.86^{+2.71}_{-2.99})\times10^{-30}$ 
& $(1.86^{+3.95}_{-4.61})\times10^{-30}$ 
& $(1.76 \pm 2.49)\times10^{-30}$ 
& $(1.76 \pm 4.99)\times10^{-30}$ \\

\hline
\end{tabular}
\label{tab:rho_constraints}
\end{table*}

In particular, in $ow$CDM models, the distribution exhibits a positive skewness, as indicated by the discrepancy between the mean and the median, and by the asymmetric percentile-based credible intervals
This behavior is expected because $z_{o\Lambda\dagger}$ is a highly non-linear derived quantity, which develops a long tail toward large values when $\Omega_{k0}$ approaches zero from below. The same case is seen in Fig.~\ref{fig:Okh0owcdm} due to $z_{ow\dagger} \propto \sqrt{-\Omega_{\Lambda}/\Omega_{k0}}$.
Therefore, percentile-based credible intervals provide a more faithful summary of the posterior than the Gaussian approximation. The large variation in $z_{ow\dagger}$ arises from its strong sensitivity to $\Omega_{k0}$ near spatial flatness. 
Since $z_{o\Lambda\dagger} \propto \sqrt{-\Omega_{\Lambda}/\Omega_{k0}}$, even small posterior weight near $\Omega_{k0}\to 0^{-}$ generates a long positive tail in the distribution. 
As a result, the posterior of $z_{ow\dagger}$ becomes highly skewed and non-Gaussian, with the mean and standard deviation strongly affected by rare samples close to flatness. 
Percentile-based credible intervals therefore provide a more robust summary of the constraint than the Gaussian approximation.

\begin{figure}[t!]
\includegraphics[width=0.5\textwidth]{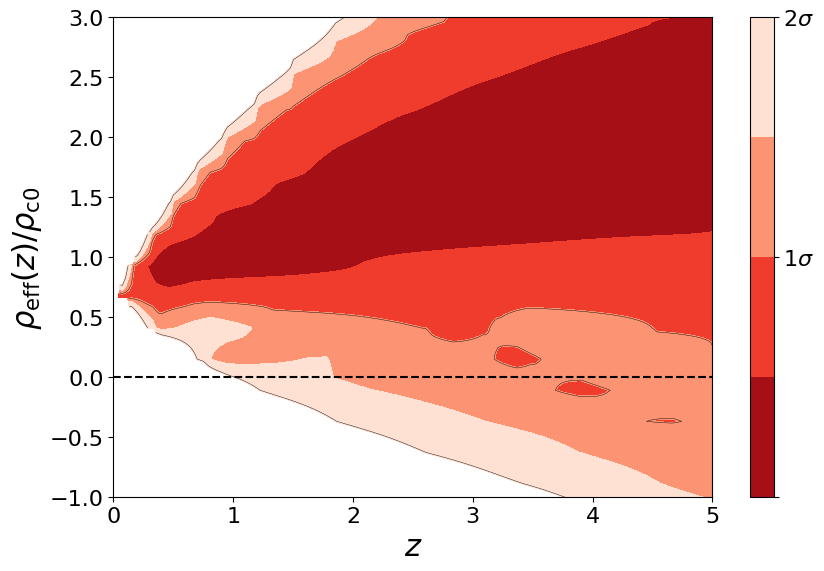}
  \caption{Redshift evolution of the effective dark-sector energy density normalized to the
present critical density, $\rho_{\rm eff}(z)/\rho_{c0}$, where
$\rho_{\rm eff}=\rho_{\rm ci}+\rho_k$ includes the contributions from the
cosmic fluid and spatial curvature. The shaded regions show the
$1\sigma$ and $2\sigma$ credible intervals obtained from the MCMC analysis
using BAO+CC+SN+SH0ES dataset combination.
The horizontal dashed line marks $\rho_{\rm eff}=0$, indicating the
possible redshift range where the effective dark-sector energy density
crosses zero.}
  \label{fig:rhotransit}
\end{figure}

\begin{figure}[t!]
\includegraphics[width=0.5\textwidth]{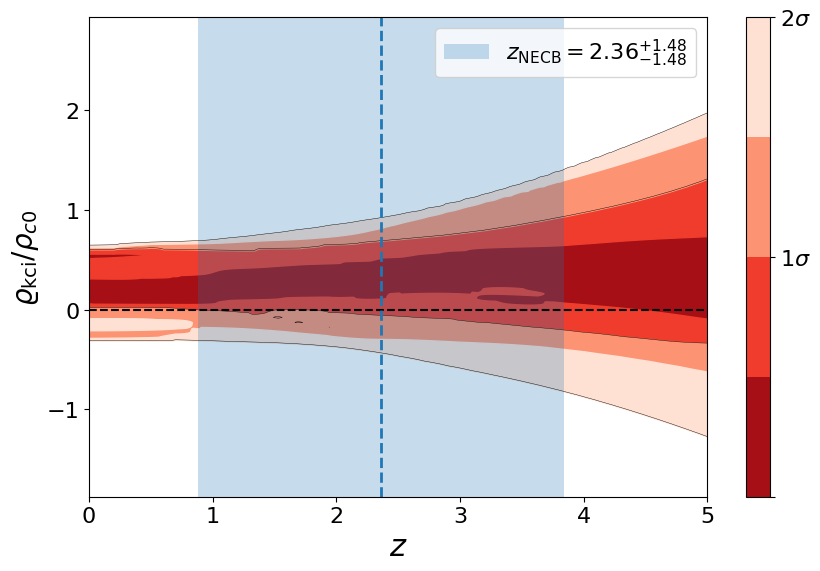}
  \caption{Redshift evolution of $\varrho_{\rm kci}/\rho_{c0}$.
Shaded regions show the $1\sigma$ and $2\sigma$ credible intervals from the
BAO+CC+SN+SH0ES data set. The horizontal dashed line denotes the NECB $\varrho_{\rm kci}=0$.
The vertical dashed line and the blue band indicate the median and
$1\sigma$ posterior interval of the transition redshift
$z_{\rm NECB}=2.36^{+1.48}_{-1.48}$.}
  \label{fig:NECB}
\end{figure}

We have additionally examined whether the percentile-based method is required for the posterior distributions of $\Omega_{\rm ci0}$ and $\varrho$. 
For the derived parameter $\varrho$, the posterior distribution is mildly non-Gaussian, so the mean and median values do not perfectly coincide (see Table~\ref{tab:rho_constraints}). 

\begin{figure}[t!]
\includegraphics[width=0.38\textwidth]{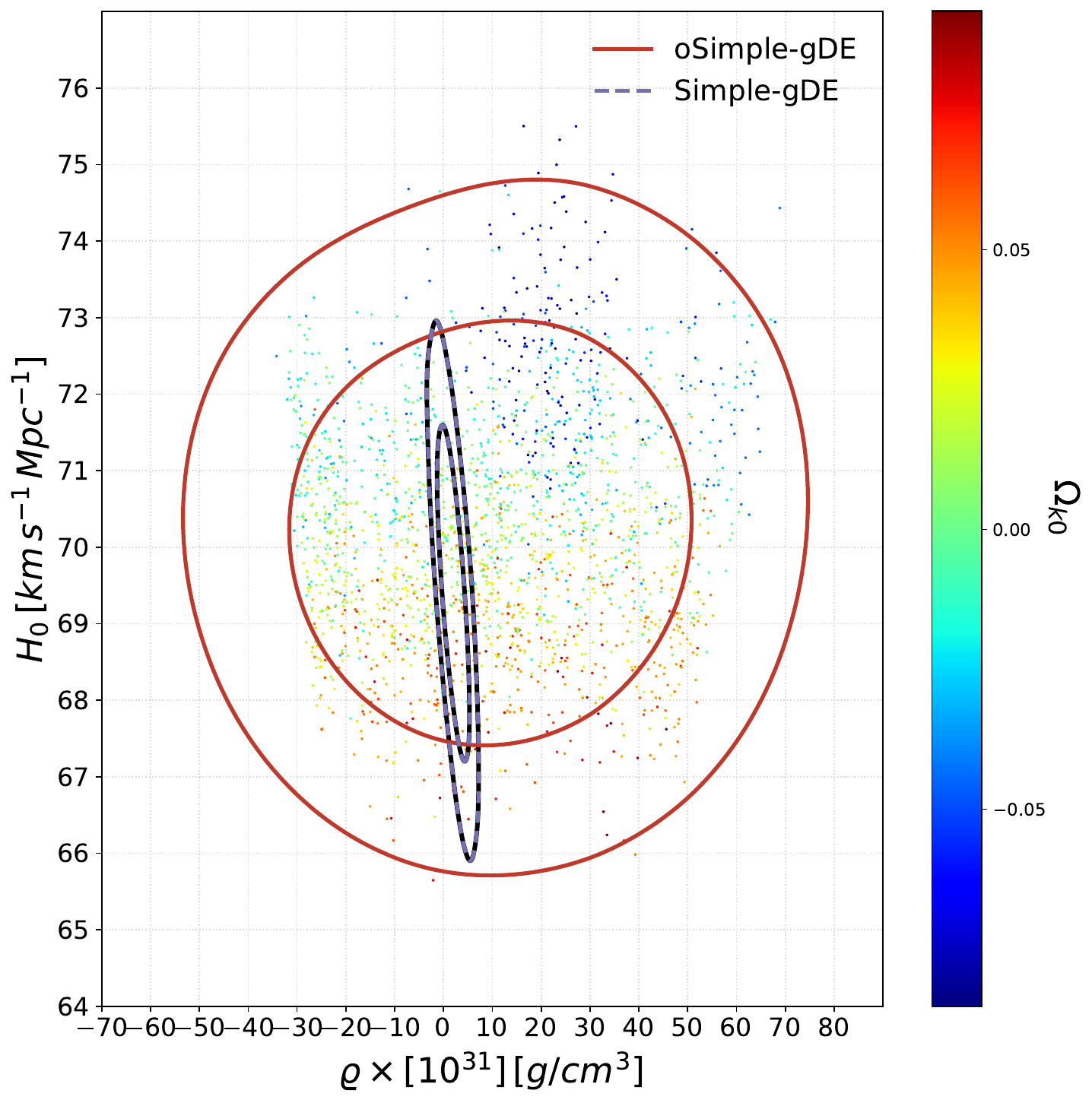}
  \caption{Constraints in the $(\varrho,H_0)$ plane for the combined BAO+CC+SN+SH0ES dataset. Contours show the $1\sigma$ and $2\sigma$ confidence regions. The colored points indicate MCMC samples, with the color scale representing the curvature parameter $\Omega_{k0}$.
}
  \label{fig:varrhoH0}
\end{figure}

For this reason, we also compute percentile-based Bayesian credible intervals and report them in Tables~\ref{tab:yesknoSh0es} and~\ref{tab:yeskyesSh0es}. Conversely, the posterior distribution for the parameter $\Omega_{\rm ci0}$ is nearly Gaussian. The percentile-based constraints, $\Omega_{\rm ci0}=0.654^{+0.006}_{-0.006}\,(68\%~\mathrm{CL})$ and $\Omega_{\rm ci0}=0.654^{+0.011}_{-0.012}\,(95\%~\mathrm{CL})$, are in excellent agreement with the corresponding Gaussian estimates ($0.654\pm0.006$) derived from the mean and standard deviation. Therefore, we report $\Omega_{\rm ci0}$ in Gaussian form (mean $\pm\sigma$ and mean $\pm2\sigma$) throughout the tables.

The rationale for using percentile-based intervals for the derived parameter $\varrho$ is that its posterior extends into negative values. This behavior is physically expected: since $\varrho \propto (1+w)\,\Omega_{\rm ci0}$, the sign of $\varrho$ is determined by the joint posterior distributions of $(1+w)$ and $\Omega_{\rm ci0}$. Because the MCMC chains allow $(1+w)$ to cross zero within the explored parameter space, $\varrho$ naturally explores both positive and negative values (see Fig.~\ref{fig:varrhoH0}). Thus, the percentile intervals reflect the genuine posterior support of the model rather than being an artifact of the statistical procedure.

Figure~\ref{fig:varrhoH0} displays the constraints in the $(\varrho,H_0)$ plane for the BAO+CC+SN+SH0ES dataset, revealing a clear degeneracy between the inertial mass density parameter $\varrho$ and the Hubble constant $H_0$. Specifically, larger values of $H_0$ tend to correlate with positive curvature contributions, as indicated by the color-coded MCMC samples. Allowing spatial curvature significantly broadens the permitted parameter space relative to the spatially flat Simple-gDE case, demonstrating that curvature can partially compensate for the effects of the inertial mass density in shaping the late-time expansion history.

\section{Conclusion}
\label{sec:final}
The statistical indistinguishability among these three models---dark energy with \textit{constant energy density}, dark energy with a \textit{constant equation-of-state parameter}, and dark energy with \textit{constant inertial mass density}---suggests that current data sets do not provide clear guidance as to which quantity should be regarded as the fundamental constant characterizing dark energy \cite{Acquaviva:2021jov}. 

For the Simple-gDE model, the Bayesian evidence differences relative to $\Lambda$CDM are $\Delta\ln\mathcal{Z}=-0.05$ and $-0.78$ for the BAO+CC+SN and BAO+CMB+SN data sets, respectively, indicating no statistically significant preference between the models. Notably, for the BAO+CC+SN data combination the cosmological constant case is not contained within the $1\sigma$ confidence region. A similar behavior is observed for the $w$CDM model, which yields $\Delta\ln\mathcal{Z}=0.44$ and $-0.48$ relative to $\Lambda$CDM for the BAO+CC+SN and BAO+CMB+SN data sets, respectively. Since the inferred IMD remains positive in all cases, none of these models exhibit a sign transition in the dark-energy density, and therefore no improvement in the cosmological tensions is obtained within these spatially flat scenarios.

On the other hand, our analysis highlights  recurring patterns that appear in many studies: \textbf{(i)} a transition from negative to positive dark-energy density in the past, typically around $z \sim 2\!-\!4$, which may be interpreted as an AdS-to-dS transition \cite{Akarsu:2021fol}, inspired or conjectured from \cite{Akarsu:2019hmw}, \textbf{(ii)} This behavior is accompanied by a later transition (as $z\rightarrow0$) of the inertial mass density, $(\rho+p)$, from negative to positive values near $z \sim 0.5$, corresponding to crossing the boundary of the Null Energy Condition, $(\rho+p=0)$ \cite{Ozulker:2025ehg,Caldwell:2025inn,Gokcen:2026pkq}. In this context, labeling dark energy simply as \textit{phantom} or \textit{quintessence}, or describing the event as a phantom divide line (PDL) crossing, may be misleading, since the energy density $\rho$ is not necessarily always positive, as first pointed out in \cite{Adil:2023exv}. Instead, the quantity $\rho + p$ may provide a more informative indicator of the physical character of dark energy.

Theoretical consistency is governed by the total cosmic fluid. As long as the Null Energy Condition is not violated by the total fluid, neither negative IMD nor negative $\rho$ necessarily implies inconsistency \cite{Caldwell:2025inn} at cosmological scales. The momentum conservation (Euler) equation, $\rho_{\text{I}}\frac{{\rm d}v^j}{{\rm d}t} = -\partial_j p$ is typically overlooked due to homogeneity assumptions. Beyond relativistic dynamics, even in the non-relativistic Newtonian regime, the pressure and the gravitational self-potential,  kinematic deviations, can contribute to the effective inertial mass density. While these effects may intertwine in determining the effective mass, the present study focuses specifically on the role of inertial mass density in the context of dark energy. The implications of this possibility should therefore be discussed and constrained both on galactic and cosmological scales.

We independently show that with some geometric/metric related modifications such as spatial curvature, we can see these recurring patterns (sign transitions in $\rho$ and $\rho+p$). The observational picture changes qualitatively once spatial curvature is allowed. For the BAO+CC+SN+SH0ES data set, the $o$Simple-gDE model yields $h_0=0.717\pm0.014$, $w_{\rm ci0}=-0.536\pm0.228$, and $\Omega_{k0}=-0.035\pm0.026$, corresponding to a positive effective inertial mass density today, $\varrho_{\rm kci}=(1.86^{+2.71}_{-2.99})\times10^{-30}\,\mathrm{g\,cm^{-3}}$, together with a physically relevant sign transition in the effective dark-energy density at
$z_{\dagger}=1.51^{+0.68}_{-0.34}$ ($1.17 \le z_{\dagger} \le 2.19$) for the $W_{-1}$ branch. The NECB transition is obtained at 
$z_{\rm NECB}=2.36^{+1.48}_{-1.48}$ 
($0.88 \le z_{\rm NECB} \le 3.84$). Data-driven preference for an NECB-crossing remains persistent. In this case, the Bayesian evidence difference $\Delta\ln\mathcal{Z}=-3.63$ provides strong support for $o$Simple-gDE relative to $o\Lambda$CDM. By contrast, for the same data set the $ow$CDM model gives only a moderate-significance transition around $z_{\dagger}=4.49^{+4.25}_{-1.63}$ and remains disfavored with $\Delta\ln\mathcal{Z}=1.65$ relative to $o\Lambda$CDM. 

The opposite sign of the inertial mass density to that of the curvature term (corresponding to a closed spatial geometry) leads to sign transitions in the past not only for the inertial mass density $\rho_{\rm DE}+p_{\rm DE}$ but also for the dark energy density $\rho_{\rm DE}$ itself. The transition in $\rho_{\rm DE}+p_{\rm DE}$ may  arise as a consequence of a sign transition in the effective dark-energy density. 
A consistent transition from a phantom regime to a quintessence-like regime cannot generally be achieved within a single canonical scalar-field framework \cite{Akarsu:2026anp}. Even remaining purely in the phantom regime is known to suffer from theoretical stability problems. Contrariwise, in the present framework, the change in the sign of $\rho+p$ arises effectively from the interplay between spatial curvature and the dark-energy sector, without invoking a fundamental phantom degree of freedom. Consequently, the usual instability problems associated with phantom models do not arise in this scenario. Construction of sign transition models from effective DE energy density via geometric extensions, spatial curvature, extra dimensions, expansion anisotropies etc. has been discussed in \cite{Katirci:2025ucs}, yet inertial mass density has not been analyzed and should be studied in this context in following studies.  

After a comparison between \emph{single-parameter models characterized by constant defining properties} in this paper, we plan to investigate updating constraints in a subsequent comparative study of two-parameter extension models based on the inertial mass density, Graduated-dark energy model and CPL parametrization.

\begin{acknowledgments}
This work is dedicated to the memory of Professor İlber Ortaylı. We are grateful to \" Ozg\" ur Akarsu and Eric Linder for useful remarks during the course of this work. L.A.E.\ acknowledges support from T\"{U}B\.{I}TAK through a postdoctoral researcher fellowship associated with Grant No.~124N627. This article/publication is based upon work from COST Action CA21136 Addressing observational tensions in cosmology with systematics and fundamental physics (CosmoVerse) supported by COST (European Cooperation in Science and Technology).
\end{acknowledgments}

\bibliographystyle{apsrev4-2_mod}
\bibliography{references}

\end{document}